\documentclass[reqno]{amsart}

\usepackage{bezier,amssymb,longtable,bbm,amsfonts,verbatim,url}

\newtheorem{theorem}{Theorem}[section]

\newtheorem{prop}[theorem]{Proposition}

\DeclareMathSymbol{\N}{\mathbin}{AMSb}{"4E}
\DeclareMathSymbol{\Z}{\mathbin}{AMSb}{"5A}
\DeclareMathSymbol{\R}{\mathbin}{AMSb}{"52}
\DeclareMathSymbol{\C}{\mathbin}{AMSb}{"43}

\newcommand{\lieg}{\mbox{${\mathfrak g}\/$}}
\newcommand{\lieh}{\mbox{${\mathfrak h}\/$}}
\newcommand{\compm}{\mbox{${\mathfrak m}\/$}}

\newlength{\defaultparindent}
\begin{document}

\title[The Petrov Classification]{Three Dimensional Lorentz homogeneous spaces and the Petrov classification}
\author[Adam Bowers]{Adam Bowers \\ \\  University of Missouri\\Department of Mathematics\\Columbia, MO 65211 USA}
\thanks{{\em Author's current e-mail address}: \url{abowers@ucsd.edu}}
\address{University of Missouri\\Department of Mathematics\\Columbia, MO 65211 USA}
\email{abowers@ucsd.edu}
\curraddr{University of California at San Diego\\Department of Mathematics\\La Jolla, CA 92093-0112 USA}

\begin{abstract}
In this note, we verify the classification of local geometries given by A.Z. Petrov.  First, we determine criteria for identifying a given 3D Lorentz homogeneous space in Petrov's classification.  Then, we identify all inequivalent 1D subalgebras of all real 4D Lie algebras and determine which of these give rise to a homogeneous space admitting an invariant Lorentz metric.
\end{abstract}


\maketitle

\newpage

\section{Introduction}

There has long been interest in the classification of Lie algebras of vector fields.  In \cite{olver-R2}, for example, Gonzalez-Lopez, Kamran, and Olver gave a classification of all real Lie field systems on $\R^2$ that generalize the original classification of Lie algebras of vector fields on $\C$ given by Lie.  
The classification of all Lie algebras of vector fields on $\R^3$ is deemed an intractable problem, and so in order to make further progress, one constructs restricted classes of Lie algebras, for instance those which leave invariant some geometric structure of the underlying manifold.

In 1961, Alexei Zinovievich Petrov gave a classification of local geometries according to their local isometry group \cite{petrov}.   Petrov's work continues to generate interest today \cite{Galaev, Haesen1, Kiosak, Marvan} (to name just a few).

With just a few notable exceptions (Petrov numbers 32.26 and 33.55) all the actions listed by Petrov are locally simple actions, and so the bulk of Petrov's classification can be viewed as the classification of homogeneous spaces with invariant Lorentz metrics.  The study of Lorentz homogeneous spaces has always received much attention, and recently this is especially true in dimension three; e.g. \cite{Calvaruso5, Calvaruso4, Calvaruso6, Haouari}.
Petrov himself described the two- and three-dimensional cases as ``classes [that] correspond to physically significant gravitational fields'' \cite[p. 211]{petrov}.

As Petrov's work can be viewed as the classification Lorentz homogeneous spaces, the discussion can be largely reduced to a purely algebraic viewpoint involving Lie algebras.
(Petrov's classification is primarily one of the classification of Lorentz metrics with symmetry, a slightly different problem than the classification of group actions admitting invariant Lorentz metrics.  For example, the metrics numbered 32.03 and 32.19 in Petrov have the same group action, however the Lorentz geometry of the manifolds is different.)

While Petrov's work represents a milestone in the classification of Lie algebras of vector fields, a number of outstanding problems remain.  For example, if a Lorentz manifold with symmetry group of dimension greater than or equal to two is given, there arises the problem of identifying the given metric with one in Petrov's list.  To illustrate the difficulty, consider the work of M. MacCallum.  In \cite{mac}, MacCallum states that Petrov ``omitted'' two metrics, which he calls A2c and B2b.  Using the results of this paper, it is possible to show that A2c can be identified with Petrov number 32.24 ($-$) and B2b with Petrov number 32.07.

In this paper, we implement the algebraic classification of three-dimensional Lorentz homogeneous spaces.  This will lead to an independent verification of Petrov's results and a concrete method for identification.  In particular, we prove the following theorem:
\begin{theorem}\label{thm-petrov}
Let $M$ be a homogeneous space of dimension $3$ with Lorentz metric $\gamma$.  If $G$ is the group of symmetries of $\gamma$ and $H$ is any isotropy subgroup of $G$, then the identification of $M$ in the Petrov classification is uniquely determined by the Lie algebra of $G$, the isotropy type (rotation, boost, or null), and the complement type (symmetric or ideal). (See Section~\ref{sec:petrov} for terminology.)
\end{theorem}

\section{Background}

At the foundation of the study of homogeneous spaces is the following well-known theorem (e.g.,  \cite[Theorem~9.3 ]{boothby}, or  \cite[Theorem~3.62]{warner}):
\begin{theorem}
\label{thm:M-quotient}
Let $G$ be a Lie group acting transitively on a manifold $M$ by a smooth action $\mu$ (i.e., $M$ is a homogeneous space). If $x$ is any point in $M$, then the mapping $F:G/G_x\rightarrow M$ defined by $F(gH)=\mu_{x}(g)$ for all $g\in~G$ is a smooth $G$-equivariant diffeomorphism.
\end{theorem}
In the above theorem, the set $G_x$ is called the {\em isotropy subgroup\/} of $G$ at $x\in M$ and is defined to be the set  $G_x = \{ g \in G \ | \ \mu(g,x) = x \}$.

To direct our study of Lorentz homogeneous spaces, an important result is taken from the literature (e.g., \cite[Section~8.5]{exact-solutions}):
\begin{theorem}
\label{thm:dimension-of-G}
Let $M$ be a manifold of dimension $n$ with metric $\gamma$. The group $G$ of symmetries of $\gamma$ is a Lie group and $\dim(G) \leq \frac{1}{2}n(n+1).$  In the case of equality, $M$ is a space of constant curvature and $G_x$ is the whole of the generalized orthogonal group for each $x\in M$.
\end{theorem}

Of primary interest in this paper is the case when $\dim(M)=3$.  In this case, the only possible non-degenerate metrics at a point (up to congruence) are given by the diagonal matrices diag$(1,1,1)$ and diag$(1,1,-1)$. These are the Riemannian and Lorentz metrics, respectively.  The orthogonal groups which preserve these metrics are O(3) for the Riemannian and O(2,1) for the Lorentz.  Theorem~\ref{thm:dimension-of-G} implies $\dim(G)\leq 6$.  Depending on the dimension of $G$, the isotropy subgroup $G_x$ can have only one of four possible dimensions:

(1) If $\dim(G) = 6$, then $\dim(G_x) = 3$.  In this case, $G_x$ must be isomorphic to either O(3) or O(2,1). The metric $\gamma$ is of constant curvature and a complete classification of Lorentz metrics is well-known \cite[pages 46-47]{koba72}.

(2) If $\dim(G) = 5$, then $\dim(G_x) = 2$.  Petrov includes in his classification of local geometries only one five-dimensional group action on three-dimensional orbits (33.01).  This action, however, is not transitive, and consequently does not lead to a homogeneous space.  (For more on this case, see \cite{mylorentz}.) 

(3) If $\dim(G) = 4$, then $\dim(G_x) = 1$.  This case is quite rich and makes up the bulk of this paper.

(4) If $\dim(G) = 3$, then $\dim(G_x) = 0$.  In this case, the manifold is diffeomorphic to the group and there is a six parameter family of $G$-invariant Lorentz metrics.

It remains then to study the case of a four-dimensional Lie group acting transitively on a three-dimensional manifold.  This reduces to an algebraic problem, by means of the following significant theorem \cite[Proposition~3.1]{koba-nomi}:

\begin{theorem}
Let $G$ be a Lie group with connected subgroup $H$, and let $G$ and $H$ have Lie algebras $\lieg$ and $\lieh$, respectively.  There is a one-to-one correspondence between $G$-invariant Lorentz metrics on $G/H$ and inner products on $\lieg/\lieh$ that are {\rm ad}$\lieh$-invariant (i.e., invariant under the action of $\lieh$ by the adjoint map).
\end{theorem}
We should remark that the inner products mentioned in the conclusion of the previous theorem need not be positive definite.  Also, in the statement of the theorem, the adjoint map ${\rm ad}:\lieg\times\lieg\rightarrow\lieg$ is given by the formula ${\rm ad}(X,Y) = [X,Y],$ where $[~,~]$ denotes the Lie bracket on $\lieg$
(e.g.,  \cite[Section~3.46]{warner}).

By Theorem~\ref{thm:M-quotient}, in order to study homogeneous spaces, it is necessary to consider only Lie groups and their isotropy subgroups.  There are generally many such isotropy subgroups, but the next theorem limits the number of quotient spaces that need to be considered.
\begin{theorem}
\label{thm:equivalent-subgroups}
Let $G$ be a Lie group with closed subgroups $H$ and $K$.  If $\phi: G \rightarrow G$ is a smooth group isomorphism such that $\phi(H) = K$, then $\phi$ induces a $G$-equivariant diffeomorphism $\tilde{\phi} : G/H \rightarrow G/K$.
\end{theorem}

To summarize, {\em in order to find an invariant metric $\gamma$ on a manifold $M$ it is enough to find an invariant inner product $\eta$ on a vector space $\lieg/\lieh$.}  In this paper, we will determine the real Lie algebra pairs $(\lieg,\lieh)$ (up to equivalence by Lie algebra automorphism) that admit an invariant inner product $\eta$ on a vector space $\lieg/\lieh$.  We will concern ourselves primarily with the case where $\dim(\lieg) = 4$ and $\dim(\lieh)=1$.  In this case, we will identify the invariant inner products with the $G$-invariant metrics in Petrov's list.

In Section~\ref{sec:petrov}, we will investigate each of the Lie algebras of infinitesimal generators given by Petrov.  The invariant information given in Tables~\ref{tbl:invariants-G3} and~\ref{tbl:invariants-G4} will be used as a means of discrimination and identification.

In Section~\ref{sec:worksheets}, we identify all inequivalent one-dimensional subalgebras $\lieh$ of each three- and four-dimensional real Lie algebra $\lieg$.  We then determine which real Lie algebra pairs $(\lieg,\lieh)$ admit an invariant inner product $\eta$ on the vector space $\lieg/\lieh$.   All the ad{$\lieh$}-actions which admit an invariant inner product on $\lieg/\lieh$ are identified with a homogeneous space from the minimal list of Petrov group actions.  We will use the classification of low dimensional real Lie algebras introduced by P. Winternitz (see Appendix~\ref{app:winternitz_tables}).

The data given in Section~\ref{sec:worksheets} will verify that Petrov's list is complete for $4$-dimensional groups acting transitively on $3$-dimensional manifolds.

%
%

\section{Local Group Actions in Petrov}
\label{sec:petrov}

In the book {\em Einstein Spaces} \cite{petrov}, the Russian mathematician and physicist Alexei Zinovievich Petrov gave a complete classification of local geometries according to their local isometry group.  In so doing, he gave a complete list of vector fields on $\R^4$ that preserve a Lorentz metric.  He constructed the vector fields as infinitesimal generators of local group actions, $\{X_1,X_2,\ldots,X_n\}$, where $n$ is the dimension of the isometry group.  In this paper, we are primarily concerned with the case $n=4$.  Petrov's work on this topic can be found in Chapter~5, Section~32 of \cite{petrov}.  We will focus on Petrov's work on orbit spaces of 3 dimensions, which include the metrics Petrov numbered 32.3 - 32.27.  We will also include the simpler case of 3 dimensional group actions on 2-dimensional orbit spaces, which include the metrics numbered 30.1 - 30.8.  (We will see in Section~\ref{sec:worksheets} that Petrov missed one in this case.)  These cases are worth investigation, for (in Petrov's words), ``these classes correspond to physically significant gravitational fields'' \cite[p. 211]{petrov}.

Some of the numbers Petrov gave correspond to a family of metrics and a family of local group actions.  These metrics are 32.11, 32.14, 32.16, 32.23, 32.24, 32.25, and 32.27.  With the exception of 32.11, the family of metrics is parameterized.  A list of these parameteriztions is given below.
$$
\begin{array}{ccccc}
32.14 & c\neq~2 & \hspace{2cm}& 32.24 & e_2=\pm~1 \\
32.16 & q^2<4 & & 32.25 & \varepsilon=0,1\\
32.23 & e_2=\pm~1 & & 32.27 & e=\pm~1
\end{array}
$$

The family of metrics numbered 32.11 is not given as a parameterization, however.  Using Petrov's notation, 32.11 has the following infinitesimal generators:  $$ X_1 = p_2, \; X_2 = p_3, \; X_3 = -p_1 \; X_4 = {x^2}{p_3} \pm {x^3}{p_2},$$  
where $(x^1,x^2,x^3,x^4)$ denotes an arbitrary point in $\R^4$, and $p_i = \frac{\partial}{\partial x^i}$,  $i=1,\ldots,4$.
In this paper, the case where $X_4 = {x^2}{p_3} + {x^3}{p_2}$ will be denoted by 32.11(+), and the case where $X_4 = {x^2}{p_3} - {x^3}{p_2}$ will be denoted by 32.11(--).  In addition, for simplicity, the metric 32.23 with $e_2 = +1$ will be denoted by 32.23(+) and 32.23 with $e_2 = -1$ will be denoted by 32.23(--). Similarly, 32.24 with $e_2 = \pm 1$ will be denoted by 32.24(+) or 32.24(--), depending on which choice of parameter is used.

Each Lie algebra of vector fields in Petrov's list can be thought of as the model for a Lorentz homogeneous space.  This identification, however, is blind to certain geometric properties, so identical group actions will lead to identical Lorentz homogeneous spaces.  Listed in Table~\ref{tbl:petrov-equiv} are the equivalent group actions.  For $n=4$, only the metrics on the left side of each column of Table~\ref{tbl:petrov-equiv} will be studied, as the other choices would yield identical results. (We will see this explicitly for metrics 30.2 and 30.8, and also for 30.4 and 30.5.)

\begin{table}[h]

\caption{\label{tbl:petrov-equiv} Local Group Action Equivalences in Petrov}

\begin{center}

\begin{tabular}{|l|l|} \hline

& \\

\begin{minipage}{2.0in}
\sloppy

30.2 \dotfill 30.8 \\
30.4 \dotfill 30.5 \\
32.03 \dotfill 32.19 \\
32.04 \dotfill 32.20 \\
32.05 \dotfill 32.21 \\
32.06 \dotfill 32.22 \\
32.07 \dotfill 32.23 $(e_2 = -1)$ \\
32.08 \\
32.09 \dotfill 32.25 $(\varepsilon = 0)$ \\
32.10 \dotfill 32.25 $(\varepsilon = 1)$ \\

\end{minipage}

&

\begin{minipage}{2.0in}
\sloppy

32.11(+) \dotfill 32.27 $(e = -1)$ \\
32.11(--) \dotfill 32.27 $(e = +1)$ \\
32.12 \\
32.14 \\
32.15 \\
32.16 \\
32.23 $(e_2 = +1)$ \\
32.24 $(e_2 = +1)$ \\
32.24 $(e_2 = -1)$ \\
32.26 \\

\end{minipage}

\\ \hline

\end{tabular}
\end{center}
\end{table}

The local group actions given by Petrov (that we will consider) are listed in Table~\ref{tbl:petrov-G3} for $n=3$ and Table~\ref{tbl:petrov-G4} for $n=4$.  Five typos in Petrov's work have been corrected in these tables.  These typos occur in numbers 30.6 (one), 30.8 (one), 32.07 (one), and 32.24 (two), and are given in Table~\ref{petrov_typos}, using Petrov's notation (as described above).

\begin{table}[hat]

\caption{\label{petrov_typos} Corrections of Typos in Petrov}

\begin{center}

\begin{tabular}{|ll|} \hline
30.6 & \\

Typo: & $X_3 = \sin{x^2}{p_1} - \cos{x^2}\tan{x^1}{p_1}$ \\

Correction: & $X_3 = \sin{x^2}{p_1} - \cos{x^2}\tan{x^1}{p_2}$ \\ \hline

30.8 & \\

Typo: & $X_3 = {x^2}{p_1} + {x^4}{p_2}$ \\

Correction: & $X_3 = {x^2}{p_1} + {x^1}{p_2}$ \\ \hline

32.07 & \\

Typo: & $X_1 = \cos{x^3}{p_2} + (\coth{x^2}\sin{x^3}-1){p_3}$ \\

Correction: & $X_1 = \cos{x^3}{p_2} - (\coth{x^2}\sin{x^3}-1){p_3}$ \\ \hline

32.24 & \\

Typo: & $X_3 = -e^{x^2}p_1 + ({x^2}^2 + e_2e^{2{x^3}})p_2 + 2{x^3}p_3$ \\

Correction: & $X_3 = -e^{x^3}p_1 + ({x^2}^2 + e_2e^{2{x^3}})p_2 + 2{x^2}p_3$ \\  \hline

\end{tabular}
\end{center}
\end{table}

For 30.8, the vector fields listed by Petrov do form a Lie algebra (which is L(3,1) in the Winternitz classification), but the group action does not admit an invariant Lorentz metric.

In the text below 32.07 (page 229), Petrov states that there is another metric which can be ``obtained by replacing all the trigonometrical functions in (32.7) by the corresponding hyperbolic functions.''  As stated this is not correct, as the new vector fields are not closed under the bracket operation.  If the vector fields are adjusted slightly, they do form a Lie algebra, but the action is identical to one already in the list.

For each Petrov group action listed in Tables~\ref{tbl:petrov-G3} and~\ref{tbl:petrov-G4}, we discern several properties and list them in the tables below.  In Tables~\ref{tbl:petrov-classify-G3} and ~\ref{tbl:petrov-classify-G4}, the Lie algebra of the infinitesimal generators is classified according to the Winternitz classification (see Appendix~\ref{app:winternitz_tables}), and a change of basis is given (setting $e_i$ equal to the $i^{\rm th}$ entry listed).  In Tables~\ref{tbl:petrov-generic-isotropy-G3} and~\ref{tbl:petrov-generic-isotropy-G4}, a basis for the isotropy $\lieh$ of each group action is given at a generic point $x = (x^1,x^2,x^3,x^4)$ in the  manifold on which the group is acting.  

The group action is transitive on the orbits, and so any point on the orbit will suffice for the calculations.  Consequently, in Tables~\ref{tbl:petrov-preferred-isotropy-G3} and~\ref{tbl:petrov-preferred-isotropy-G4}, a ``preferred'' point $x_0$ is chosen so that the calculations will be simple.  The isotropy at the preferred point $x_0$ is denoted by $\lieh_0$.

If $\lieh$ is a Lie subalgebra of a Lie algebra $\lieg$, then a {\em reductive complement\/}  is a subspace $\compm$ of $\lieg$ such that $\lieg = \lieh \oplus \compm$ (vector space direct sum), where  $[\lieh, \compm]\subseteq\compm.$  In this case, the Lie algebra pair ($\lieg$, $\lieh$) is said to be a {\em reductive Lie algebra pair}.  For the cases we consider, a reductive complement always exists (e.g., \cite[Theorem~2.1]{fels-renner}) and is listed under ``Reductive Complement of $\lieh_0$.''

The existence of a reductive complement simplifies computations by virtue of the following proposition, given here without proof:
\begin{prop}
If the Lie algebra pair $(\lieg, \lieh)$ admits a reductive complement $\compm$, then there exists a vector space isomorphism $\psi:\lieg/\lieh\rightarrow\compm$ which is invariant under the ${\rm ad}\lieh$ action.
\end{prop}

In Tables~\ref{tbl:invariants-G3} and~\ref{tbl:invariants-G4}, the classification data is summarized.  For each Petrov number, we indicate the Lie algebra according to the Winternitz classification and we list the preferred isotropy $\lieh_0$ in the Winternitz basis (given by the change of basis formulas in Tables~\ref{tbl:petrov-classify-G3} and ~\ref{tbl:petrov-classify-G4}).

The reductive complements $\compm$ in Tables~\ref{tbl:petrov-preferred-isotropy-G3} and~\ref{tbl:petrov-preferred-isotropy-G4} often are either symmetric ($[\compm, \compm]\subseteq\lieh_0$) or an ideal ($[\lieg, \compm]\subseteq \compm $).   When the complement $\compm$ is either symmetric or an ideal (or both), a note is made to that effect under ``Complement Type.''  We use $S$ to indicate a symmetric complement and $I$ to indicate an ideal.

The last entry in Tables~\ref{tbl:invariants-G3} and~\ref{tbl:invariants-G4} contains the ``Isotropy Type.''  The possible isotropy types are {\em rotations}, {\em boosts}, and {\em nulls}.  The type of isotropy can be determined by the eigenvalues of the linear isotropy representation matrix, $M$, of the map ${\rm ad}(h_0)|_{\compm}$. (Here $h_0$ is a vector in the one-dimensional subspace $\lieh_0$).  If $M$ has two real non-zero eigenvalues, the isotropy type is a {\em boost}.  If $M$ has two imaginary non-zero eigenvalues, the isotropy type is a {\em rotation}.  Finally, if all the eigenvalues of $M$ are zero, then the isotropy type is {\em null}.  We use the notation $B$, $R$, and $N$ for boost, rotation, and null, respectively.

%
%

\subsection{Three-dimensional group actions: $G_3$ on $V_2$.}

In this section, we consider Petrov's $3$-dimensional group actions with $2$-dimensional orbits, or in Petrov's terminology $G_3$ on $V_2$ (or $V_2^\ast$).  For Petrov's work, see \cite[pp. 206 - 210]{petrov}.

In Table~\ref{tbl:petrov-G3}, we list the vector fields given by Petrov.  They are given at a point $x=(x^1, x^2, x^3)$ in $\R^3$.  We use the symbol $\partial_{x^k}$ to denote the partial derivative with respect to $x^k$, where $k\in\{1,2,3\}$.

\renewcommand{\arraystretch}{1.4}

\begin{longtable}{|ll|}

\caption{\label{tbl:petrov-G3} Petrov Vector Fields ($G_3$ on $V_2$)} \\ \hline
\endfirsthead
\caption{{\em Continued.}}\\ \hline
\endhead
\hline \multicolumn{2}{r}{{\em Continued on next page.}}
\endfoot
\endlastfoot

30.1 \hspace{6ex} &
$
X_1 = \partial_{x^1}, \;\; 
X_2 = \partial_{x^2}, \;\;
X_3 = -x^2\partial_{x^1}+{x^1}\partial_{x^2}. \;\;
$ \\ \hline

30.2 &
$
X_1 = \partial_{x^1}, \;\; 
X_2 = \partial_{x^2}, \;\; 
X_3 = x^1\partial_{x^1}-{x^2}\partial_{x^2}. \;\; 
$ \\ \hline

30.4 &
$
X_1 = \cosh(x^2)\partial_{x^1} + \sinh(x^2)\,\tan(x^1)\partial_{x^2}, \;\; 
X_2 = \partial_{x^2}, \;\; 
$ \\

& $X_3 = \sinh(x^2)\partial_{x^1} + \cosh(x^2)\,\tan(x^1)\partial_{x^2}. \;\; 
$ \\ \hline

30.5 &
$
X_1 = \cos(x^2)\partial_{x^1} - \sin(x^2)\,\tanh(x^1)\partial_{x^2}, \;\; 
X_2 = \partial_{x^2}, \;\; 
$ \\

& $X_3 = \sin(x^2)\partial_{x^1} + \cos(x^2)\,\tanh(x^1)\partial_{x^2}. \;\; 
$ \\ \hline

30.6 &
$
X_1 = \cos(x^2)\partial_{x^1} + \sin(x^2)\,\tan(x^1)\partial_{x^2}, \;\; 
X_2 = \partial_{x^2}, \;\; 
$ \\

& $X_3 = \sin(x^2)\partial_{x^1} - \cos(x^2)\,\tan(x^1)\partial_{x^2}. \;\; 
$ \\ \hline

30.8 &
$
X_1 = \partial_{x^1}, \;\; 
X_2 = \partial_{x^2}, \;\;
X_3 = x^2\partial_{x^1}+{x^1}\partial_{x^2}. \;\;
$ \\ \hline

\end{longtable}

In Table~\ref{tbl:petrov-classify-G3}, we classify the Lie algebra generated by the vector fields listed in Table~\ref{tbl:petrov-G3}.  We use the classification of Winternitz (see Appendix~\ref{app:winternitz_tables}).  The change of basis is given by taking $e_i$ to be the $i^{\rm th}$ entry in the ``Change of Basis'' column.

\renewcommand{\arraystretch}{1.5}

\begin{longtable}{|l|l|l|}
\caption{\label{tbl:petrov-classify-G3} 3-D Gravitational Fields on 2-D Orbits: $G_3$ on $V_2$.}
\endfirsthead
\caption{{\em Continued.}}
\endhead
\multicolumn{2}{r}{{\em Continued on next page.}}
\endfoot
\endlastfoot
\hline
{\em Petrov Number} & {\em Winternitz Classification} & {\em Change of Basis} \\ \hline
30.1 & L(3,4,$x=0$) & $X_1, X_2, -X_3$  \\ \hline
30.2 & L(3,2,$x=-1$) & $X_1, X_2, X_3$  \\ \hline
30.4 & L(3,5) & $X_1-X_3, X_2, -X_1-X_3$  \\ \hline
30.5 & L(3,5) & $X_2-X_3, X_1, -X_2-X_3$  \\ \hline
30.6 & L(3,6) & $-X_3, X_2, X_1$ \\ \hline
30.8 & L(3,2,$x=-1$) & $X_1+X_2, X_1-X_2, X_3$ \\ \hline
\end{longtable}

In Table~\ref{tbl:petrov-generic-isotropy-G3}, we list a basis for the (one-dimensional) isotropy $\lieh$ for the group actions described by the vector fields in Table~\ref{tbl:petrov-G3}.  The isotropy is given at a generic point $x=(x^1, x^2, x^3)$.

\

\renewcommand{\arraystretch}{1.75}
\begin{longtable}{|l|l|}
\caption{\label{tbl:petrov-generic-isotropy-G3} Generic Isotropy for Local Group Actions: $G_3$ on $V_2$}
\endfirsthead
\caption{{\em Continued.}}
\endhead
\multicolumn{2}{r}{{\em Continued on next page.}}
\endfoot
\endlastfoot
\hline
{\em Petrov Number} & {\em Generic Isotropy $\lieh$ at $x=(x^1, x^2, x^3, x^4)$} \\ \hline
30.1 & $x^2X_1-x^1X_2+X_3$  \\ \hline
30.2 & $-x^1X_1+x^2X_2+X_3$  \\ \hline
30.4 & $-\sinh(x^2)X_1-\tan(x^1)X_2+\cosh(x^2)X_3$  \\ \hline
30.5 & $-\sin(x^2)X_1-\tanh(x^1)X_2+\cos(x^2)X_3$  \\ \hline
30.6 & $-\sin(x^2)X_1+\tan(x^1)X_2+\cos(x^2)X_3$ \\ \hline
30.8 & $-x^2X_1-x^1X_2+X_3$ \\ \hline
\end{longtable}

In Table~\ref{tbl:petrov-preferred-isotropy-G3}, we choose a ``preferred'' point $x_0$ and evaluate $\lieh$ at that point.  The resulting isotropy is denoted $\lieh_0$.  We also exhibit a reductive complement for $\lieh_0$.  (It is the same for each example.)

\begin{longtable}{|l|c|c|c|} 
\caption{Preferred Isotropy for Local Group Actions: $G_3$ on $V_2$}
\label{tbl:petrov-preferred-isotropy-G3} 
\endfirsthead
\endhead
\endfoot
\endlastfoot
\hline
{{\em Petrov Number}} & {\em $x_0$} & {\em Isotropy $\lieh_0$ at $x_0$} & {\em Reductive Complement}  \\ \hline
30.1 -- 30.8 & $x_0=(0,0,0)$ & $X_3$ & $X_1, X_2$ \\ \hline
\end{longtable}

In Table~\ref{tbl:invariants-G3}, we give the isotropy $\lieh_0$ in the corresponding Winternitz basis, using the change of basis listed in Table~\ref{tbl:petrov-classify-G3} (with a possible change of sign).  We then list the complement type and isotropy type, as described in the beginning of this section.

\begin{longtable}{|l|l|c|c|c|} 
\caption{Classification Summary: $G_3$ on $V_2$}
\label{tbl:invariants-G3} 
\endfirsthead
\endhead
\endfoot
\endlastfoot

\hline
{{\em Petrov \#}}  & {\em Winternitz Class} & $\lieh_0$ & {\em Complement Type} & {\em Isotropy Type} \\ \hline
30.1 & L(3, 4, $x=0$) & $e_3$ & S, I & R \\ \hline
30.2 & L(3, 2, $x=-1$) & $e_3$ & S, I & B \\ \hline
30.4 & L(3, 5) & $\frac{1}{2}(e_1+e_3)$ & S & B \\ \hline
30.5 & L(3, 5) & $\frac{1}{2}(e_1+e_3)$ & S & B \\ \hline
30.6 & L(3, 6) & $e_1$ & S & R \\ \hline
30.8 & L(3, 2, $x=-1$) & $e_3$ & S, I & B \\ \hline
\end{longtable}


\subsection{Four-dimensional group actions.}

In this section, we consider $4$-dimensional group actions with $3$-dimensional orbits, or in Petrov's terminology $G_4$ on $V_3$ (or $V_3^\ast$).  For Petrov's complete list, see \cite[pp. 228 - 233]{petrov}.

In Table~\ref{tbl:petrov-G4}, we list the {\em inequivalent} vector fields given by Petrov.  They are given at a point $x=(x^1, x^2, x^3, x^4)$ in $\R^4$.  As before, we use the symbol $\partial_{x^k}$ to denote the partial derivative with respect to $x^k$, where now $k\in\{1,2,3,4\}$.

In Table~\ref{tbl:petrov-classify-G4}, we classify the Lie algebra generated by the vector fields listed in Table~\ref{tbl:petrov-G4}.  As before, we use the classification in Appendix~\ref{app:winternitz_tables}.  The change of basis is given by taking $e_i$ to be the $i^{\rm th}$ entry in the ``Change of Basis'' column.

In Table~\ref{tbl:petrov-generic-isotropy-G4}, we list a basis for the (one-dimensional) isotropy $\lieh$ for the group actions described by the vector fields in Table~\ref{tbl:petrov-G4}.  The isotropy is given at a generic point $x=(x^1, x^2, x^3, x^4)$.

In Table~\ref{tbl:petrov-preferred-isotropy-G4}, we choose a ``preferred'' point $x_0$ and evaluate $\lieh$ at that point.  The resulting isotropy is denoted $\lieh_0$.  We also exhibit a reductive complement for $\lieh_0$. ({\em Note:} The group action in 32.26 is {\em not transitive}, and consequently the isotropy at a generic point does not conjugate to an isotropy at a preferred point.)

In Table~\ref{tbl:invariants-G4}, we give a basis for the isotropy $\lieh_0$ in the Winternitz basis, using the change of basis listed in Table~\ref{tbl:petrov-classify-G4}.  We also list the complement type ({\em Comp.}) and isotropy type ({\em Iso.}), using the notation described in the beginning of this section.

\begin{longtable}{|ll|}
\caption{\label{tbl:petrov-G4} Inequivalent Petrov Vector Fields ($G_4$ on $V_3$)} \\ \hline
\endfirsthead
\caption{{\em Continued.}}\\ \hline
\endhead
\hline \multicolumn{2}{r}{{\em Continued on next page.}}
\endfoot
\endlastfoot

32.03 \hspace{6ex} &
$
X_1 = \partial_{x^2}, \;\; 
X_2 = \partial_{x^3}, \;\;
X_3 = -\partial_{x^1}+{x^3}\partial_{x^2}, \;\;
$ \\

& $X_4 = -{x^1}\partial_{x^1}+{x^3}\partial_{x^3}$ \\ \hline

32.04 &
$
X_1 = \partial_{x^2}, \;\; 
X_2 = \partial_{x^3}, \;\; 
X_3 = -\partial_{x^1}+{x^3}\partial_{x^2}, \;\; 
$ \\

& $X_4 = -{x^3}\partial_{x^1}+\frac{1}{2}((x^3)^2-(x^1)^2)\partial_{x^2}+{x^1}\partial_{x^3}$ \\ \hline

32.05 &
$
X_1 = \partial_{x^2}, \;\; 
X_2 = \partial_{x^3}, \;\; 
X_3 = -\partial_{x^1}+{x^2}\partial_{x^2}+{x^3}\partial_{x^3}, \;\; 
$ \\

& $X_4 = -{x^3}\partial_{x^2}+{x^2}\partial_{x^3}$ \\ \hline

32.06 &
$
X_1 = \partial_{x^2}, \;\; 
X_2 = \partial_{x^3}, \;\;
X_3 = -\partial_{x^1}+{x^2}\partial_{x^2}+{x^3}\partial_{x^3}, \;\;
$ \\

& $X_4 = -{x^3}\partial_{x^2}+{x^2}\partial_{x^3}$ \\ \hline

32.07 
& $X_1 = \cos(x^3)\partial_{x^2} + \frac{-\cosh(x^2)\sin(x^3)+\sinh(x^2)}{\sinh(x^2)}\partial_{x^3}, \;\;$ \\
& $X_2 = \sin(x^3)\partial_{x^2} + \frac{\cosh(x^2)\cos(x^3)}{\sinh(x^2)}\partial_{x^3}, \;\;$ \\
& $X_3 = \cos(x^3)\partial_{x^2}-\frac{\cosh(x^2)\sin(x^3)+\sinh(x^2)}{\sinh(x^2)}\partial_{x^3}, \;\;$ \\
& $X_4 = \partial_{x^1}$ \\ \hline

32.08 &
$X_1 = e^{-x^3}\partial_{x^1}-e^{-x^3}{(x^2)^2}\partial_{x^2}-2x^2e^{-x^3}\partial_{x^3}, \;\;$ \\

&
$ 
X_2 = \partial_{x^3}, \;\; 
X_3 = e^{x^3}\partial_{x^2}, \;\; 
X_4 = \partial_{x^1}
$ \\ \hline

32.09 
&$X_1 = \partial_{x^3}, \;\;$ \\
& $X_2 = \sin(x^3)\partial_{x^2} + \frac{\cos(x^2)\cos(x^3)}{\sin(x^2)}\partial_{x^3}, \;\;$ \\
& $X_3 = \cos(x^3)\partial_{x^2} -\frac{\cos(x^2)\sin(x^3)}{\sin(x^2)}\partial_{x^3}, \;\;$ \\
& $X_4 = \partial_{x^1}$ \\ \hline

32.10 &
$X_1 = \partial_{x^2}, \;\;$ \\

 & $X_2 = \cos(x^2)\partial_{x^1}-\frac{\cos(x^1)\sin(x^2)}{\sin(x^1)}\partial_{x^2}+\frac{\sin(x^2)}{\sin(x^1)}\partial_{x^3}, \;\;$ \\

 & $X_3 = -\sin(x^2)\partial_{x^1}-\frac{\cos(x^1)\cos(x^2)}{\sin(x^1)}\partial_{x^2}+\frac{\cos(x^2)}{\sin(x^1)}\partial_{x^3}, \;\;$ \\
 
 & $X_4 = \sin(x^3)\partial_{x^1}-\frac{\cos(x^3)}{\sin(x^1)}\partial_{x^2}+\frac{\cos(x^1)\cos(x^3)}{\sin(x^1)}\partial_{x^3}$ \\ \hline

32.11 (+) &
$
X_1 = \partial_{x^2}, \;\; 
X_2 = \partial_{x^3}, \;\;
X_3 = -\partial_{x^1}, \;\;
X_4 = {x^3}\partial_{x^2}+{x^2}\partial_{x^3} 
$  \\ \hline

32.11 (--) &
$
X_1 = \partial_{x^2}, \;\; 
X_2 = \partial_{x^3}, \;\;
X_3 = -\partial_{x^1}, \;\;
X_4 = {-x^3}\partial_{x^2}+{x^2}\partial_{x^3} 
$ \\ \hline

32.12 &
$
X_1 = \partial_{x^2}, \;\; 
X_2 = \partial_{x^3}, \;\;
X_3 = -\partial_{x^1}, \;\;
X_4 = -{x^3}\partial_{x^1}+{x^2}\partial_{x^3} 
$ \\ \hline

32.14 $(c \neq 2)$ &
$
X_1 = \partial_{x^2}, \;\; 
X_2 = \partial_{x^3}, \;\;
X_3 = {x^3}\partial_{x^2}-{x^1}\partial_{x^3}, \;\;
$ \\

& $X_4 = (2-c){x^1}\partial_{x^1}+c{x^2}\partial_{x^2}+{x^3}\partial_{x^3}$ \\ \hline

32.15 &
$
X_1 = \partial_{x^2}, \;\; 
X_2 = \partial_{x^3}, \;\;
X_3 = {x^3}\partial_{x^2}-{x^1}\partial_{x^3}, \;\;
$ \\

& $ X_4 = \partial_{x^1}+2{x^2}\partial_{x^2}+{x^3}\partial_{x^3}$ \\ \hline

32.16 $(q^2 < 4)$ &
$
X_1 = \partial_{x^2}, \;\; 
X_2 = \partial_{x^3}, \;\;
X_3 = {x^3}\partial_{x^2}-{x^1}\partial_{x^3}, \;\;
$ \\

& $X_4 = -((x^1)^2+qx^1+1)\partial_{x^1}+(qx^2+\frac{1}{2}(x^3)^2)\partial_{x^2}-{x^1x^3}\partial_{x^3}$ \\ \hline

32.23 (+) &
$
X_1 = \partial_{x^2}, \;\; 
X_2 = {x^2}\partial_{x^2}+\partial_{x^3}, \;\;
$ \\

&
$
X_3 = ((x^2)^2+e^{2x^3})\partial_{x^2}+2{x^2}\partial_{x^3}, \;\;
X_4 = \partial_{x^1}
$ \\ \hline

32.24 $(\epsilon = \pm 1)$ &
$
X_1 = \partial_{x^2}, \;\; 
X_2 = {x^2}\partial_{x^2}+\partial_{x^3}, \;\;
$ \\

& $X_3 = -e^{x^3}\partial_{x_1} + ((x^2)^2+{\epsilon}e^{2x^3})\partial_{x^2} + 2{x^2}\partial_{x^3}, \;\;$ \\

& $X_4 = \partial_{x^1}$ \\ \hline

32.26 &
$
X_1 = \partial_{x^1}, \;\; 
X_2 = \partial_{x^2}, \;\;
X_3 = \partial_{x^3}, \;\;
$ \\

& $X_4 = {x^2}\partial_{x_1} + {\omega(x^4)}\partial_{x^2} + {\lambda(x^4)}\partial_{x^3}$ \\ \hline

\end{longtable}

\begin{longtable}{|l|l|l|}
\caption{\label{tbl:petrov-classify-G4} 4-D Gravitational Fields on 3-D Orbits: $G_4$ on $V_3$.}
\endfirsthead
\caption{{\em Continued.}}
\endhead
\multicolumn{3}{r}{{\em Continued on next page.}}
\endfoot
\endlastfoot

\hline

{\em Petrov Number} & {\em Winternitz Class} & {\em Change of Basis} \\ \hline

32.03 & L(4,8) & $X_1, X_2, X_3, X_4$ \\ \hline

32.04 & L(4,11) & $X_1, X_2, X_3, -X_4$ \\ \hline

32.05 & L(4,$-2$) & $X_1, X_3, X_2, X_4$ \\ \hline

32.06 & L(4,13) & $X_1, X_2, X_3, -X_4$ \\ \hline

32.07 & L(4,$-7$) & $X_1, X_2, X_3, X_4$ \\ \hline

32.08 & L(4,$-7$) & $-X_1, X_2, X_3, X_4$ \\ \hline

32.09 & L(4,$-8$) & $-X_3, X_2, X_1, X_4$ \\ \hline

32.10 & L(4,$-8$) & $X_1, X_2, X_3, X_4$ \\ \hline

32.11 (+)  & L(4,$-4$, $x=-1$) & $X_1+X_2, -X_1+X_2, X_4, X_3$ \\ \hline

32.11 (--) & L(4,$-6$, $x=0$) & $X_1, X_2, -X_4, X_3$ \\ \hline

32.12 & L(4,1) & $X_3, X_2, X_1, X_4$ \\ \hline

32.14 ($c \neq 0, c \neq 1$) & L(4,9,$x=c-1$) & $X_1, X_2, X_3, X_4$ \\ \hline

32.14 ($c=1$) & L(4,7) & $X_1, X_2, X_3, X_4$ \\ \hline

32.14 ($c=0$) & L(4,8) & $X_1, X_2, X_3, X_4$ \\ \hline

32.15 & L(4,10) & $X_1, X_2, X_3, X_4$ \\ \hline

32.16 ($q \neq 0$) & L$\left(4,12,x=\sqrt{\frac{q^2}{4-q^2}}\right)$  & $-\frac{4-q^2}{2q}xX_1, \ \ -\frac{q}{2}X_2+X_3,$ \\
 & & \phantom{aaaa}  $-\frac{4-q^2}{2q}xX_2, \  \  \frac{2}{q}xX_4$  \\ \hline

32.16 ($q = 0$)    & L(4,11) & $X_1, X_3, -X_2, -X_4$ \\ \hline

32.23 (+) & L(4,$-7$) & $-X_1, X_2, X_3, X_4$ \\ \hline

32.24 (+) & L(4,$-7$) & $-X_1, X_2, X_3, X_4$ \\ \hline

32.24 (--) & L(4,$-7$) & $X_3, -X_2, -X_1, X_4$ \\ \hline

32.26  & L(4,$-3$) & $X_1, X_2, X_4, X_3$ \\ \hline

\end{longtable}


\renewcommand{\arraystretch}{1.75}
\begin{longtable}{|l|l|}
\caption{\label{tbl:petrov-generic-isotropy-G4} Generic Isotropy for Local Group Actions: $G_4$ on $V_3$}
\endfirsthead
\caption{{\em Continued.}}
\endhead
\multicolumn{2}{r}{{\em Continued on next page.}}
\endfoot
\endlastfoot
\hline
{\em Petrov Number} & {\em Generic Isotropy $\lieh$ at $x=(x^1, x^2, x^3, x^4)$} \hspace{3cm} \\ \hline
32.03 & $x^1x^3X_1-x^3X_2-x^1X_3+X_4$ \\ \hline
32.04 & $\frac{1}{2}\big((x^1)^2+(x^3)^2\big)X_1-x^1X_2-x^3X_3+X_4$ \\ \hline
32.05 & $x^2X_1-x^3X_2-X_3+X_4$ \\ \hline
53.06 & $x^3X_1-x^2X_2+X_4$ \\ \hline

32.07 & $\frac{1}{2}\big( \sin(x^3)\tanh(x^2)+1 \big) X_1 - \cos(x^3)\tanh(x^2) X_2$ \\
           & \hspace{1cm} $+ \frac{1}{2}\big( \sin(x^3)\tanh(x^2)-1 \big) X_3$ \\ \hline

32.08 & $X_1+ 2x^2 e^{-x^3} X_2 + (x^2)^2 e^{-2x^3} X_3 - e^{-x^3}X_4$ \\ \hline

32.09 & $-\cot(x^2) X_1 + \cos(x^3) X_2  - \sin(x^3) X_3$ \\ \hline

32.10 & $\cos(x^3) \sin(x^1) X_1$ \\
           & \hspace{1cm} $- \big(\sin(x^2) \cos(x^1) \cos(x^3) + \sin(x^3) \cos(x^2)\big) X_2$ \\
           & \hspace{1cm} $+ \big(\sin(x^2) \sin(x^3) - \cos(x^1) \cos(x^3) \cos(x^2)\big) X_3  + X_4$ \\ \hline
32.11 (+) & $-x^3X_1-x^2X_2+X_4$ \\ \hline
32.11 ($-$) & $x^3X_1-x^2X_2+X_4$ \\ \hline
32.12 & $-x^2X_2-x^3X_3+X_4$ \\ \hline
32.14 ($c \neq 0, 1$) & $-x^3X_1+x^1X_2+X_3$ \\ \hline
32.14 ($c = 1$) & $-x^3X_1+x^1X_2+X_3$ \\ \hline
32.14 ($c = 0$) & $-x^3X_1+x^1X_2+X_3$ \\ \hline
32.15 & $-x^3X_1+x^1X_2+X_3$ \\ \hline
32.16 ($q \neq 0$) & $-x^3X_1+x^1X_2+X_3$ \\ \hline
32.16 ($q = 0$) & $-x^3X_1+x^1X_2+X_3$ \\ \hline
32.23 (+) & $\big((x^2)^2-\exp(2x^3)\big)X_1-2x^2X_2+X_3$ \\ \hline
32.24 (+) & $\big((x^2)^2-\exp(2x^3)\big)X_1-2x^2X_2+X_3+\exp(x^3)X_4$ \\ \hline
32.24 ($-$) & $\big((x^2)^2+\exp(2x^3)\big)X_1-2x^2X_2+X_3+\exp(x^3)X_4$ \\ \hline
32.26 & $-x^2X_1-\omega(x^4)X_2-\lambda(x^4)X_3+X_4$ \\ \hline
\end{longtable}


\begin{longtable}{|l|c|c|c|} 
\caption{Preferred Isotropy for Local Group Actions: $G_4$ on $V_3$}
\label{tbl:petrov-preferred-isotropy-G4} 
\endfirsthead
\endhead
\endfoot
\endlastfoot
\hline
{{\em Petrov Number}} & {\em $x_0$} & {\em Isotropy $\lieh_0$ at $x_0$} & {\em Reductive Complement of $\lieh_0$}  \\ \hline
32.03 & (0,0,0,0) & $X_4$ & $X_1, -X_2+X_3, X_2+X_3$  \\ \hline
32.04 & (0,0,0,0) & $X_4$ & $X_1, X_2, X_3$  \\ \hline
32.05 & (0,0,0,0) & $-X_3+X_4$ & $X_3, X_1+X_2, X_1-X_2$  \\ \hline
32.06 & (0,0,0,0) & $X_4$ & $X_3, X_1,X_2$   \\ \hline
32.07 & $(0,0,0,0)$ & $\frac{1}{2}(X_1-X_3)$ & $X_4, X_2, -\frac{1}{2}(X_1+X_3)$   \\ \hline
32.08 & (0,0,0,0) & $X_1 - X_4$ & $\frac{1}{2}X_3, -X_2, -X_1$   \\ \hline
32.09 & (0,$\frac{\pi}{2}$,$\frac{\pi}{2}$,0) & $-X_3$ & $X_4, X_2, X_1$ \\ \hline
32.10 & ($\frac{\pi}{2}$,0,0,0) & $X_1 + X_4$ & $X_1, X_3, X_2$  \\ \hline
32.11($+$) & (0,0,0,0) & $X_4$ & $X_3, X_1, -X_2$   \\ \hline
32.11($-$) & (0,0,0,0) & $X_4$ & $X_3, X_1, X_2$  \\ \hline
32.12 & (0,0,0,0) & $X_4$ & $-X_1, X_2, X_3$  \\ \hline
32.14 ($c \neq 0, 1$)  & (1,0,0,0) & $X_2+X_3$ & $\frac{1}{2-c}X_4, \frac{1}{2-c}X_2+\frac{c-1}{2-c}X_3, X_1$  \\ \hline
32.14 ($c = 1$) & (1,0,0,0) & $X_2+X_3$ & $X_4, X_2, X_1$  \\ \hline
32.14 ($c = 0$) & (1,0,0,0) & $X_2+X_3$ & $X_4, X_2-X_3, 2X_1$  \\ \hline
32.15 & (0,0,0,0) & $X_3$ & $X_4, X_2+X_3, X_1$ \\ \hline
32.16 ($q \neq 0$) & ($-\frac{q}{2}$,0,0,0) & $-\frac{q}{2}X_2+X_3$ & $-X_4, X_2-qX_3, X_1$  \\ \hline
32.16 ($q = 0$) & (0,0,0,0) & $X_3$ & $-X_4, X_2, X_1$  \\ \hline
32.23($+$) & (0,0,0,0) & $\frac{1}{2}(X_1-X_3)$ & $X_4, \frac{1}{2}(X_1+X_3), X_2$  \\ \hline
32.24($+$) & (0,0,0,0) & $\frac{1}{2}(X_1-X_3-X_4)$ & $X_1-X_3, 2X_2, X_1+X_3$  \\ \hline
32.24($-$) & (0,0,0,0) & $\frac{1}{2}(X_1+X_3+X_4)$ & $X_1+X_3, -2X_2, X_1-X_3$  \\ \hline
32.26 & N/A &  &    \\ \hline
\end{longtable}

\newpage

\begin{longtable}{|l|c|c|c|c|} 
\caption{Classification Summary: $G_4$ on $V_3$}
\label{tbl:invariants-G4} 
\endfirsthead
\endhead
\endfoot
\endlastfoot
\hline
{{\em Petrov \#}} & Winternitz Class & $\lieh_0$  & {\em Comp.} & {\em Iso.} \\ \hline
32.03 & L(4, 8) & $e_4$ & I & B \\ \hline
32.04 & L(4, 11) & $e_4$ & I & R \\ \hline
32.05 & L(4, $-2$) & $e_2-e_4$ &  I & B \\ \hline
32.06 & L(4, 13) & $e_4$ &  I & R  \\ \hline
32.07 & L(4, $-7$) & $\frac{1}{2}(e_1-e_3)$ &  S & R \\ \hline
32.08 & L(4, $-7$) & $e_1+e_4$   & I & N \\ \hline
32.09 & L(4, $-8$) & $e_1$ & S & R \\ \hline
32.10 & L(4, $-8$) & $e_1 + e_4$  & I & R \\ \hline
32.11($+$) & L(4, $-4$, $x=-1$) & $e_3$ & S, I & B \\ \hline
32.11($-$) &  L(4, $-6$, $x=0$) & $e_3$ &  S, I & R \\ \hline
32.12 & L(4,1) & $e_4$ & S, I & N \\ \hline
32.14 ($c \neq 0, 1$) & L(4, 9, $x=c-1$)  & $e_2+e_3$  &  & N \\ \hline
32.14 ($c = 1$) & L(4,7) & $e_2+e_3$  & I & N \\ \hline
32.14 ($c = 0$) & L(4,8) & $e_2+e_3$  & S & N \\ \hline
32.15 & L(4,10) & $e_3$ &  & N \\ \hline
32.16 ($q \neq 0$) &  L$\left(4, 12, x=\sqrt{\frac{q^2}{4-q^2}}\right)$& $e_2$ & & N \\ \hline
32.16 ($q = 0$) & L(4, 11) & $e_2$ &   S & N \\ \hline
32.23($+$) &  L(4, $-7$) & $\frac{1}{2}(e_1+e_3)$  & S & B \\ \hline
32.24($+$) &  L(4, $-7$) & $\frac{1}{2}(e_1+e_3+e_4)$  & I & B \\ \hline
32.24($-$) &   L(4, $-7$) & $\frac{1}{2}(e_1-e_3+e_4)$  & I & R \\ \hline
32.26 & L(4, $-3$)  & N/A  &  &  \\ \hline
\end{longtable}


\section{Computation of invariant metrics from Lie algebras with one-dimensional isotropy}
\label{sec:worksheets}

\subsection{Preamble to Calculations}
\label{sec:preamble}

In the following sections, we calculate the inequivalent (up to Lie algebra automorphism) one-dimensional Lie subalgebras $\lieh$ of a given Lie algebra $\lieg$, and then determine if the Lie algebra pairs ($\lieg$,$\lieh$) admit a Lorentz (or Riemannian) metric.  Each section is divided as follows:

The first item listed for each Lie algebra is an array that summarizes the data accumulated for each Lie algebra.  The first column in the array contains the {\em inequivalent} (up to automorphism) one-dimensional Lie subalgebras $\lieh$ of the given Lie algebra $\lieg$.  The second column contains a complement $\compm$ of that Lie subalgebra.  
The column ``Complement Type'' indicates if the complement $\compm$ is reductive ({\em R\/}), symmetric ({\em S\/}), or an ideal ({\em I\/}).  Recall that in order for a complement to be either symmetric or an ideal, it must be reductive.

The fourth column is labeled $\kappa$.  This column lists the dimension of the nullspace of the ad action of $\lieh$ on the quotient \lieg/$\lieh$.  The value of $\kappa$ is an invariant of the automorphism group.

The fifth column lists a basis for the quadratic forms on the quotient \lieg/$\lieh$ invariant under the ad action of $\lieh$ on the quotient \lieg/$\lieh$.  The notation is explained in Table~\ref{tbl:qf-basis}.

\renewcommand{\arraystretch}{1}

\begin{table}[h]
\caption{Notation Used for Basis of Quadratic Forms \label{tbl:qf-basis}}

\begin{tabular}{|c|}\hline
\\
Lie Algebras of 3 Dimensions \\
\\
$
D_1 = \left(\begin{array}{cc} 1 & 0 \\ 0 & 0 \\ \end{array} \right), \;\;
R   = \left(\begin{array}{cc} 1 & 0 \\ 0 & 1 \\ \end{array} \right), \;\;
B   = \left(\begin{array}{cc} 1 & 0 \\ 0 & -1\\ \end{array} \right), \;\;
$ \\
\\
$
Q = \{ %
\left(\begin{array}{cc} 1 & 0 \\ 0 & 0 \\ \end{array} \right), \; %
\left(\begin{array}{cc} 0 & 0 \\ 0 & 1 \\ \end{array} \right), \; %
\left(\begin{array}{cc} 0 & 1 \\ 1 & 0 \\ \end{array} \right) \}, \;\;  %

Z = \left(\begin{array}{cc} 0 & 0\\ 0 & 0\\ \end{array} \right).
$\\
\\ 

\\
Lie Algebras of 4 Dimensions \\
\\
$
D_1 = \left(\begin{array}{cccc} 1 & 0 & 0\\ 0 & 0 & 0\\ 0 & 0 & 0\\ \end{array} \right), \;\;
D_2 = \left(\begin{array}{cccc} 0 & 0 & 0\\ 0 & 1 & 0\\ 0 & 0 & 0\\ \end{array} \right), \;\;
D_3 = \left(\begin{array}{cccc} 0 & 0 & 0\\ 0 & 0 & 0\\ 0 & 0 & 1\\ \end{array} \right),
$ \\

\\

$
Q_{12} = \left(\begin{array}{cccc} 0 & 1 & 0\\ 1 & 0 & 0\\ 0 & 0 & 0\\ \end{array} \right), \;\;
Q_{13} = \left(\begin{array}{cccc} 0 & 0 & 1\\ 0 & 0 & 0\\ 1 & 0 & 0\\ \end{array} \right), \;\;
Q_{23} = \left(\begin{array}{cccc} 0 & 0 & 0\\ 0 & 0 & 1\\ 0 & 1 & 0\\ \end{array} \right),
$ \\

\\

$
R = \left(\begin{array}{cccc} 0 & 0 & 0\\ 0 & 1 & 0\\ 0 & 0 & 1\\ \end{array} \right), \;\;
B = \left(\begin{array}{cccc} 0 & 0 & 0\\ 0 & 1 & 0\\ 0 & 0 & -1\\ \end{array} \right), \;\;
N = \left(\begin{array}{cccc} 0 & 0 & 1\\ 0 & 1 & 0\\ 1 & 0 & 0\\ \end{array} \right),
$ \\

\\

$
Q = \{D_1, D_2, D_3, Q_{12}, Q_{13}, Q_{23}\}, \;\;\;\;  Z = {\bf 0}.
$\\

\\ \hline

\end{tabular}
\end{table}

In some cases (when a restricted inner product is admitted), the Lie algebra pair ($\lieg$,$\lieh$) represents an entry in the Petrov list, and so in this event, the Petrov number is listed in the sixth column.  Several of the cases admit a non-restricted inner product (with a six-dimensional basis of quadratic forms).  In this case the ad action is not effective and so there is not a corresponding Petrov number.

Following the table of subalgebras, there is a list of the automorphisms used in identifying those inequivalent one-dimensional subalgebras.  Listed are only those automorphisms used; there are in general more than needed for the purposes of this paper.  For the inner automorphisms, the following notation is used:
$$ A_i(t) := {\rm Ad}(te_i) = e^{t \times {\rm ad}(e_i)}. $$
The outer automorphisms are labeled $O_1(t), O_2(t), \ldots, O_k(t)$, but as there is no prefered order for outer automorphisms, the labeling is insignificant.  In some cases a disconnected subgroup of automorphisms is used. These automorphisms are of the following three types: $S_i(t), S_{ij}(t)$, or $S_i^j(t)$.  If $v = y^1e_1 + y^2e_2 + \cdots + y^ne_n$ is a given vector in the $n$-dimensional Lie algebra $\lieg$, then the three above automorphisms are defined as follows:
$$
\begin{aligned}
S_i(t)(v) & := y^1e_1 + \cdots + \frac{y^i}{t}e_i + \cdots + y^ne_n, \\
S_{ij}(t)(v) & := y^1e_1 + \cdots + \frac{y^i}{t}e_i + \cdots \frac{y^j}{t}e_j \cdots + y^ne_n, \\
S_i^j(t)(v) & := y^1e_1 + \cdots + \frac{y^i}{t}e_i + \cdots (ty^j)e_j \cdots + y^ne_n.
\end{aligned}
$$
These automorphisms, said to be {\em scalings in the $i$th term\/} or {\em scalings in the $i$th and $j$th terms}, form a disconnected subgroup of automorphisms because in each case $t>0$ or $t<0$. Throughout the following pages there is a scaling function $\lambda$ which acts on the entire vector $v = y^1e_1 + \cdots + y^ne_n$:
$$ \lambda(t)(v) := \frac{v}{t} = \frac{1}{t}(y^1e_1 + \cdots + y^ne_n) = \frac{y^1}{t}e_1 + \cdots + \frac{y^n}{t}e_n. $$
In general, $\lambda$ is not a Lie algebra automorphism, but we desire to find only a convenient {\em representative} of a one-dimensional Lie subalgebra.  Thus, any scalar multiple of a transformed vector will do.  The listed automorphisms transform an arbitrary representative to a representative in some canonical form.  Following the action of the automorphisms, the $\lambda$ function is used to scale the representative to a more ``aesthetically pleasing'' vector. It is for this reason that the $\lambda$ function is always applied last.  

Also listed here are the invariants of the Adjoint action.  The Adjoint invariants are used as a starting point to help distinguish the separate cases in the ensuing calculations.  Different values for the invariants may lead to inequivalent subalgebras. Note that the invariants of the inner automorphisms need not be invariants of the outer automorphisms.

In some cases, for formatting reasons, we write $A_i$ to mean $A_i(t)(v)$, where $t\in\R$ and $v\in\lieg$.  Whenever $v$ is written, it will be understood to be $v = y^1e_1 + \cdots + y^ne_n$, where $\{e_1, \ldots, e_n\}$ is a basis for $\lieg$.

After listing the Lie algebra automorphisms to be used, we display the computations done to find the inequivalent one-dimensional subalgebras.  These computations are divided into multiple cases.  Note that in some of the cases, we make use of the {\em sign} function:
$${\rm sign}(x) := \begin{cases} +1 & \mbox{if } x \geq 0 \\ -1 & \mbox{if } x < 0. \end{cases}$$
Note that we adopt the convention ${\rm sign}(0) = 1$.

\setlength{\parindent}{0pt}

\small

\subsection{Lie Algebras of 3 Dimensions}
\label{sec:3d}

\subsubsection{L(3,1)}

$$
\begin{tabular}{|c|c|c|c|c|} \hline
{\em Subalgebra } & {\em Complement} & {\em Complement}  & {$\kappa$}      & {\em Quadratic}  \\
{\lieh}        & {\compm}      & {\em Type}         & {}           & {\em Forms}     \\ \hline
$e_1$ & $e_2 + c_1\lieh, e_3 + c_2\lieh$    &   S     &  2  &   $Q$         \\ \hline
$e_2$ & $e_3 + c\lieh, e_1$                         &   S, I  &  1  &   $D_1$      \\ \hline
\end{tabular}
$$

$$
A_2=\left( \begin{array}{c} y^1+ty^3\\ y^2\\ y^3 \end{array} \right), \;\;
A_3=\left( \begin{array}{c} y^1-ty^2\\ y^2\\ y^3 \end{array} \right), \;\;
O_1=\left( \begin{array}{c} y^1 \\ \cos(t)y^2+\sin(t)y^3\\ -\sin(t)y^2+\cos(t)y^3 \end{array} \right).
$$

The invariants of the Adjoint action are $I=y^2$ and $J=y^3$.

Case 1. \ \   $I^2+J^2 \neq 0$:   
$$\lambda\left(\frac{\sqrt{I^2+J^2}}{{\rm sign}(y^2)}\right) \circ O_1\left(\arctan\frac{y^3}{y^2}\right) \circ A_3\left(\frac{y^1y^2}{I^2+J^2}\right) \circ A_2\left(-\frac{y^1y^3}{I^2+J^2}\right)(v) = e_2. $$

Case 2. \ \   $I =0, J = 0$:  \ \ $\lambda(y^1)(v) = e_1.$

\subsubsection{L(3,2,$x$), $x \neq 1$}

$$
\begin{tabular}{|c|c|c|c|c|c|} \hline
{\em Subalgebra } & {\em Complement} & {\em Complement}  & {$\kappa$}      & {\em Quadratic} & {\em Petrov}  \\
{\lieh}        & {\compm}      & {\em Type}         & {}           & {\em Forms} & {\em Number}    \\ \hline
$e_1$ & $e_2 + c_1\lieh, e_3 + c_2\lieh$                 &            &  2  &   $Q$   &    \\ \hline
$e_2$ & $e_1 + c_1\lieh, e_3 + c_2\lieh$                 &            &  2  &   $Q$    &   \\ \hline
$e_3 (x \neq -1)$     & $e_1, e_2$                       &   S, I     &  0  &   $Z$   &    \\ \hline
$e_3 (x=-1)$          & $e_1, e_2$                       &   S, I     &  0  &   $B$   &  30.2, 30.8  \\ \hline
$e_1+e_2 (x \neq -1)$ & $e_3 + c\lieh, e_1+xe_2$         &   R        &  1  &   $D_1$  &   \\ \hline
$e_1+e_2 (x=-1)$ & $e_3 + c\lieh, e_1-e_2$               &   S        &  1  &   $D_1$  &   \\ \hline
\end{tabular}
$$

$$
A_1(t)(v)=\left( \begin{array}{c} y^1+ty^3\\ y^2\\ y^3 \end{array} \right), \;\;
A_2(t)(v)=\left( \begin{array}{c} y^1\\ y^2+txy^3\\ y^3 \end{array} \right), \;\;
S_1(t)(v)=\left( \begin{array}{c} y^1/t \\ y^2\\y^3 \end{array} \right).
$$

The invariant of the Adjoint action is $I=y^3$.

Case 1. \ \   $I \neq 0$:  \ \ $\lambda(y^3) \circ A_2(-\frac{y^2}{xy^3}) \circ A_1(-\frac{y^1}{y^3})(v) = e_3.$

Case 2. \ \  $I =0, y^1 \neq 0, y^2 \neq 0$: \ \ $\lambda(y^2) \circ S_1(\frac{y^1}{y^2})(v) = e_1 + e_2.$

Case 3. \ \  $I =0, y^1 = 0, y^2 \neq 0$:  \ \  $\lambda(y^2)(v) = e_2.$

Case 4. \ \  $I =0, y^1 \neq 0, y^2 = 0$: \ \  $\lambda(y^1)(v) = e_1.$

\subsubsection{L(3,2, $x = 1$)}

$$
\begin{tabular}{|c|c|c|c|c|} \hline
{\em Subalgebra } & {\em Complement} & {\em Complement}  & {$\kappa$}      & {\em Quadratic}  \\
{\lieh}        & {\compm}      & {\em Type}         & {}           & {\em Forms}     \\ \hline
$e_1$ & $e_2 + c_1\lieh, e_3 + c_2\lieh$         &             &  2  &   $Q$         \\ \hline
$e_3$ & $e_1, e_2$                                           &   S, I     &  0  &   $Z$          \\ \hline
\end{tabular}
$$

$$
A_1=\left( \begin{array}{c} y^1+ty^3\\ y^2\\ y^3 \end{array} \right), \;\;
A_2=\left( \begin{array}{c} y^1\\ y^2+ty^3\\ y^3 \end{array} \right), \;\;
O_1=\left( \begin{array}{c} \cos(t)y^1+\sin(t)y^2 \\ -\sin(t)y^1+\cos(t)y^2\\y^3 \end{array} \right).
$$

The invariant of the Adjoint action is $I=y^3$.

Case 1. \ \   $I \neq 0$:  \ \   $\lambda(y^3) \circ A_2(-\frac{y^2}{y^3}) \circ A_1(-\frac{y^1}{y^3})(v) = e_3.$

Case 2. \ \   $I =0$:  \ \  $\lambda\left(\frac{(y^1)^2+(y^2)^2}{{\rm sign}(y^1)}\right) \circ O_1\left(\arctan\frac{y^2}{y^1}\right)(v) = e_1.$

\subsubsection{L(3,3)}

$$
\begin{tabular}{|c|c|c|c|c|} \hline
{\em Subalgebra } & {\em Complement} & {\em Complement}  & {$\kappa$}      & {\em Quadratic}  \\
{\lieh}        & {\compm}      & {\em Type}         & {}           & {\em Forms}     \\ \hline
$e_1$ & $e_2 + c_1\lieh, e_3 + c_2\lieh$         &             &  2  &   $Q$         \\ \hline
$e_2$ & $e_3 + c\lieh, e_1 + e_2$                    &   R       &  1  &   $D_1$     \\ \hline
$e_3$ & $e_1, e_2$                                           &   S, I     &  0  &   $Z$          \\ \hline
\end{tabular}
$$

$$
A_1=\left( \begin{array}{c} y^1+ty^3\\ y^2\\ y^3 \end{array} \right), \;\;
A_2=\left( \begin{array}{c} y^1+ty^3\\ y^2+ty^3\\ y^3 \end{array} \right), \;\;
A_3=\left( \begin{array}{c} e^{-t}(y^1-ty^2) \\ e^{-t}y^2\\y^3 \end{array} \right).
$$

The invariant of the Adjoint action is $I=y^3$.

Case 1. \ \   $I \neq 0$:  \ \   $\lambda(y^3) \circ A_2(-\frac{y^2}{y^3}) \circ A_1(\frac{y^2-y^1}{y^3})(v) = e_3.$

Case 2. \ \  $I =0, y^2 \neq 0$: \ \  $\lambda\left(y^2e^{-\frac{y^1}{y^2}}\right) \circ A_3(\frac{y^1}{y^2})(v) = e_2.$

Case 3.  \  \   $I =0, y^2 = 0$:  \ \   $\lambda(y^1)(v) = e_1.$

\subsubsection{L(3,4, $x$), $x \neq 0$}

$$
\begin{tabular}{|c|c|c|c|c|} \hline
{\em Subalgebra } & {\em Complement} & {\em Complement}  & {$\kappa$}      & {\em Quadratic}  \\
{\lieh}        & {\compm}      & {\em Type}         & {}           & {\em Forms}     \\ \hline
$e_1$ & $e_3 + c\,\lieh, xe_1 - e_2$                   &   R       &  1  &   $D_1$     \\ \hline
$e_3$ & $e_1, e_2$                                           &   S, I     &  0  &   $Z$          \\ \hline
\end{tabular}
$$

$$
A_1=\left( \begin{array}{c} y^1+txy^3\\ y^2-ty^3\\ y^3 \end{array} \right), \;\;
A_2=\left( \begin{array}{c} y^1+ty^3\\ y^2+txy^3\\ y^3 \end{array} \right), \;\;
A_3=\left( \begin{array}{c} e^{-tx}(\cos(t)y^1-\sin(t)y^2) \\ e^{-tx}(\sin(t)y1+\cos(t)y^2)\\y^3 \end{array} \right).
$$

The invariant of the Adjoint action is $I=y^3$.

Case 1. \ \  $I \neq 0$: \ \  $\lambda(y^3) \circ A_2\left(-\frac{y^1+xy^2}{y^3(x^2+1)}\right) \circ A_1\left(\frac{y^2-xy^1}{y^3(x^2+1)}\right)(v) = e_3.$

Case 2. \ \  $I =0$:  \ \  $\lambda\left(\frac{\sqrt{(y^1)^2+(y^2)^2}}{{\rm sign}(y^1)}e^{x\left(\arctan\frac{y^2}{y^1}\right)}\right) \circ A_3\left(-\arctan\frac{y^2}{y^1}\right)(v) = e_1.$

\subsubsection{L(3,4, $x=0$)}

$$
\begin{tabular}{|c|c|c|c|c|c|} \hline
{\em Subalgebra } & {\em Complement} & {\em Complement}  & {$\kappa$}      & {\em Quadratic}  & {\em Petrov} \\
{\lieh}        & {\compm}      & {\em Type}         & {}           & {\em Forms}  & {\em Number}   \\ \hline
$e_1$ & $e_3 + c\, \lieh,  -e_2$                    &   S       &  1  &   $D_1$ &    \\ \hline
$e_3$ & $e_2, e_1$                                   &   S, I     &  0  &   $R$    &  30.1     \\ \hline
\end{tabular}
$$

$$
A_1=\left( \begin{array}{c} y^1\\ y^2-ty^3\\ y^3 \end{array} \right), \;\;
A_2=\left( \begin{array}{c} y^1+ty^3\\ y^2\\ y^3 \end{array} \right), \;\;
A_3=\left( \begin{array}{c} \cos(t)y^1-\sin(t)y^2 \\ \sin(t)y1+\cos(t)y^2\\y^3 \end{array} \right).
$$

The invariant of the Adjoint action is $I=y^3$.

Case 1. \ \  $I \neq 0$:  \ \  $\lambda(y^3) \circ A_2(-\frac{y^1}{y^3}) \circ A_1(\frac{y^2}{y^3})(v) = e_3.$

Case 2. \ \  $I =0$:    \ \   $\lambda\left(\frac{\sqrt{(y^1)^2+(y^2)^2}}{{\rm sign}(y^1)}\right) \circ A_3\left(-\arctan\frac{y^2}{y^1}\right)(v) = e_1.$

\subsubsection{L(3,5)}

$$
\begin{tabular}{|c|c|c|c|c|c|} \hline
{\em Subalgebra } & {\em Complement} & {\em Complement}  & {$\kappa$}      & {\em Quadratic}  & {\em Petrov} \\
{\lieh}        & {\compm}      & {\em Type}         & {}           & {\em Forms}  & {\em Number}   \\ \hline
$e_1$      &   $e_3 + c_1\lieh, -2e_2 + c_2 \lieh$                &    & 1 & $D_1$   &     \\ \hline
$\frac{1}{2}(e_1 + e_3)$   & $\frac{1}{2}(e_1 - e_3),e_2$  & S & 0 & $B$   &   30.4, 30.5     \\ \hline
$\frac{1}{2}(e_1 - e_3)$   & $\frac{1}{2}(e_1 + e_3),e_2$  & S & 0 & $R$   &   Missing    \\ \hline
\end{tabular}
$$

$$
A_1=\left( \begin{array}{c} y^1+ty^2-t^2y^3\\ y^2-2ty^3\\ y^3 \end{array} \right), \;\;
A_3=\left( \begin{array}{c} y^1\\ 2ty^1+y^2\\ -t^2y^1-ty^2+y^3 \end{array} \right), \;\;
S_1^3=\left( \begin{array}{c} y^1/t \\ y^2 \\ ty^3 \end{array} \right).
$$

The invariant of the Adjoint action is $I=4y^1y^3+(y^2)^2$.

Case 1. \ \  $I \neq 0, y^3 \neq 0$:  
$$\lambda\left(\frac{I}{\sqrt{|I|}}\right) \circ S_1^3\left(\frac{1}{2}\frac{\sqrt{|I|}}{y^3}\right) \circ A_1\left(\frac{1}{2}\frac{y^2}{y^3}\right)(v) = \frac{1}{2}\Big(e_1 + {\rm sign}(I)\,e_3\Big).$$

Case 2. \ \  $I \neq 0, y^3=0$:  \ \ $\lambda(y^2) \circ A_3(-1) \circ A_1\left(\frac{y^2-2y^1}{2y^2}\right)(v) = \frac{1}{2}(e_1 + e_3).$

Case 3. \ \  $I=0, y^1 \neq 0$: \ \  $\lambda(y^1) \circ A_3\left(-\frac{1}{2}\frac{y^2}{y^1}\right)(v) = e_1.$

Case 4. \ \  $I=0, y^1=0$:  \ \  $\lambda(-y^3) \circ A_3(-1) \circ A_1(1)(v) = e_1.$

\subsubsection{L(3,6)}

$$
\begin{tabular}{|c|c|c|c|c|c|} \hline

{\em Subalgebra } & {\em Complement}  & {\em Complement}  & {$\kappa$}   & {\em Quadratic} & {\em Petrov}    \\
{\lieh}        & {\compm}       & {\em Type}         & {}        & {\em Forms}  & {\em Number} \\ \hline
$e_1$       & $e_3, e_2$   &    S    &   0   & $R$  & 30.6    \\ \hline
\end{tabular}
$$

$$
A_1(t)(v)=\left( \begin{array}{c} y^1\\ \cos(t)y^2-\sin(t)y^3\\ \sin(t)y^2+\cos(t)y^3 \end{array} \right), \;\;
A_3(t)(v)=\left( \begin{array}{c} \cos(t)y^1-\sin(t)y^2\\ \sin(t)y^1+\cos(t)y^2\\ y^3 \end{array} \right).
$$

The invariant of the Adjoint action is $I=(y^1)^2+(y^2)^2+(y^3)^2$.

 (Note: $I \neq 0$)
$$\lambda\left(\frac{\sqrt{I}}{{\rm sign}(y^1)}\right) \circ A_3\left(-\arctan\left(\frac{\sqrt{(y^2)^2+(y^3)^2}}{y^1{\rm sign}(y^2)}\right)\right) \circ A_1\left(-\arctan\frac{y^3}{y^2}\right)(v) = e_1.$$

\subsubsection{L(3,-1)}

$$
\begin{tabular}{|c|c|c|c|c|} \hline
{\em Subalgebra } & {\em Complement} & {\em Complement}  & {$\kappa$}      & {\em Quadratic}  \\
{\lieh}        & {\compm}      & {\em Type}         & {}           & {\em Forms}     \\ \hline
$e_1$    & $e_2 + c_1\lieh, e_3 + c_2\lieh$     &             &  2   &   $Q$             \\ \hline
$e_2$    & $e_3, e_1$                                       &   S, I     &  1  &   $D_1$          \\ \hline
$e_3$    & $e_1, e_2+c\lieh$                            &    I        &  2   &   $Q$             \\ \hline
$e_1+e_3$    & $e_2+c\lieh, e_1$                    &   I         &  1   &   $D_1$         \\ \hline
\end{tabular}
$$

$$
A_1(t)(v)=\left( \begin{array}{c} y^1+ty^2\\ y^2\\ y^3 \end{array} \right), \;\;
O_1(t)(v)=\left( \begin{array}{c} y^1\\ y^2\\ y^3+ty^2 \end{array} \right), \;\;
S_1(t)(v)=\left( \begin{array}{c} y^1/t \\ y^2\\ y^3 \end{array} \right).
$$

The invariants of the Adjoint action are $I=y^2$ and $J=y^3$.

Case 1. \ \  $I \neq 0$:  \ \   $\lambda(y^2) \circ O_1(-\frac{y^3}{y^2}) \circ A_1(-\frac{y^1}{y^2})(v) = e_2.$

Case 2. \ \  $I = 0, y^1 \neq 0, y^3 \neq 0:$  \ \   $\lambda(y^3) \circ S_1(\frac{y^1}{y^3})(v) = e_1 + e_3.$

Case 3. \ \  $I = 0, y^1 \neq 0, y^3=0:$ \ \   $\lambda(y^1)(v) = e_1.$

Case 4. \ \  $I = 0, y^1=0, y^3 \neq 0$:  \ \   $\lambda(y^3)(v) = e_3.$

\

\subsection{Lie Algebras of 4 Dimensions}
\label{sec:4d}

\subsubsection{L(4,1)}

\noindent $$
\begin{tabular}{|c|c|c|c|c|c|} \hline
{\em Subalgebra } & {\em Complement} & {\em C}  & {$\kappa$}      & {\em Quadratic}   & {\em Petrov} \\
{\lieh}        & {\compm}      & {\em Type}         & {}           & {\em Forms}        & {\em Number}  \\ \hline
$e_1$ & $e_2 + c_1\lieh, e_3 + c_2\lieh, e_4 + c_3\lieh$ & R        & 3 & $Q$                &         \\ \hline
$e_2$ & $e_4 + c\lieh, e_3, e_1$                         & R        & 2 & $D_1, D_2, Q_{12}$ &         \\ \hline
$e_3$ & $e_1, e_4 + c\lieh, e_2$                         & I        & 2 & $D_1, D_2, Q_{12}$ &         \\ \hline
$e_4$ & $-e_3, e_2, e_1$                                      & S, I     & 1 & $D_1, N$           &  32.12  \\ \hline
\end{tabular}
$$

$$
A_2=\left( \begin{array}{c} y^1+ty^4 \\ y^2\\ y^3\\ y^4 \end{array} \right), \;
A_3=\left( \begin{array}{c} y^1 \\ y^2+ty^4\\ y^3\\ y^4 \end{array} \right), \;
A_4=\left( \begin{array}{c} y^1-ty^2+\frac{1}{2}{t^2}y^3 \\ y^2-ty^3\\ y^3\\ y^4 \end{array} \right),
$$

$$
[O_1]^t=\left( \begin{array}{cccc} y^1 & y^2 & y^3+ty^4 & y^4 \end{array} \right), \;\;
[O_2]^t=\left( \begin{array}{cccc} y^1+ty^3 & y^2 & y^3 & y^4 \end{array} \right).
$$

The invariants of the Adjoint action are $I=y^3$ and $J=y^4$.

Case 1. \ \  $J \neq 0$:  \ \   $\lambda(y^4) \circ O_1(-\frac{y^3}{y^4}) \circ A_3(-\frac{y^2}{y^4}) \circ A_2(-\frac{y^1}{y^4})(v) = e_4.$

Case 2. \ \  $I \neq 0, J = 0$: \ \  $\lambda(y^3) \circ O_2\left(-\frac{y^1y^3-\frac{1}{2}(y^2)^2}{(y^3)^2}\right) \circ A_4(\frac{y^2}{y^3})(v) = e_3.$

Case 3. \ \  $I=0, J=0, y^2 \neq 0$:  \ \ $\lambda(y^2) \circ A_4(\frac{y^1}{y^2})(v) = e_2.$

Case 4. \ \  $I=0, J=0, y^2=0$: \ \  $\lambda(y^1)(v) = e_1.$

\subsubsection{L(4,2,$x$,$y$), $x \neq y$, $x \neq 1$}

$$
\begin{tabular}{|c|c|c|c|c|} \hline
{\em Subalgebra } & {\em Complement} & {\em Complement}  & {$\kappa$}      & {\em Quadratic}    \\
{\lieh}        & {\compm}      & {\em Type}         & {}           & {\em Forms}         \\ \hline
$e_1$              &  $e_2 + c_1\lieh, e_3 + c_2\lieh, e_4 + c_3\lieh$     &      & 3 &  $Q$                      \\ \hline
$e_2$              &  $e_1 + c_1\lieh, e_3 + c_2\lieh, e_4 + c_3\lieh$     &      & 3 &  $Q$                       \\ \hline
$e_3$              &  $e_1 + c_1\lieh, e_2 + c_2\lieh, e_4 + c_3\lieh$     &      & 3 &  $Q$                       \\ \hline
$e_4 (y \neq -x,-1)$              &  $e_1, e_2, e_3$                                      & S, I & 0 &  $Z$                       \\ \hline
$e_4 (y = -x)$     &  $e_1, e_2 + e_3, -e_2 + e_3$                         & S, I & 0 &  $B$                       \\ \hline
$e_4 (y = -1)$     &  $e_2, e_1 + e_3, -e_1 + e_3$                         & S, I & 0 &  $B$                       \\ \hline
$e_1 + e_2$        &  $e_4 + c_1\lieh, e_3 + c_2\lieh, e_1 + xe_2$         & R    & 2 &  $D_1, D_2, Q_{12}$        \\ \hline
$e_1 + e_3$        &  $e_4 + c_1\lieh, e_2 + c_2\lieh, e_1 + ye_3$         & R    & 2 &  $D_1, D_2, Q_{12}$        \\ \hline
$e_2 + e_3$        &  $e_4 + c_1\lieh, e_1 + c_2\lieh, xe_2 + ye_3$        & R    & 2 &  $D_1, D_2, Q_{12}$       \\ \hline
$e_1 + e_2 + e_3$  &  $e_4 + c_1\lieh, e_2 + c_2\lieh,$                    & R    & 2 &  $D_1, D_2, Q_{12}$        \\
                   &  $e_1 + xe_2 + ye_3$                                  &      &   &                           \\ \hline
\end{tabular}
$$

$$
A_1=\left( \begin{array}{c} y^1+ty^4\\ y^2\\ y^3\\ y^4 \end{array} \right), \;\;
A_2=\left( \begin{array}{c} y^1 \\ y^2+txy^4\\ y^3\\ y^4 \end{array} \right), \;\;
A_3=\left( \begin{array}{c} y^1\\ y^2\\ y^3+tyy^4\\ y^4 \end{array} \right), \;\;
$$

$$
[S_1(t)(v)]^t=\left( \begin{array}{cccc} \frac{y^1}{t} & y^2 & y^3 & y^4 \end{array} \right), \;\;
[S_2(t)(v)]^t=\left( \begin{array}{cccc} y^1 & \frac{y^2}{t} & y^3 & y^4 \end{array} \right).
$$

The invariant of the Adjoint action is $J=y^4$.

Case 1. \ \   $J \neq 0$: \ \   $\lambda(y^4) \circ A_3(-\frac{y^3}{yy^4}) \circ A_2(-\frac{y^2}{xy^4}) \circ A_1(-\frac{y^1}{y^4})(v) = e_4.$

Case 2. \ \  $J = 0, y^1 \neq 0, y^2 = 0, y^3 = 0$:  \ \  $\lambda(y^1)(v) = e_1.$

Case 3. \ \   $J = 0, y^1 \neq 0, y^2 = 0, y^3 \neq 0$: \ \  $\lambda(y^3) \circ S_1(\frac{y^1}{y^3})(v) = e_1 + e_3.$

Case 4. \ \  $J = 0, y^1 \neq 0, y^2 \neq 0, y^3 = 0$: \ \  $\lambda(y^2) \circ S_1(\frac{y^1}{y^2})(v) = e_1 + e_2.$

Case 5. \ \  $J = 0, y^1 \neq 0, y^2 \neq 0, y^3 \neq 0$: \ \  $\lambda(y^3) \circ S_2(\frac{y^2}{y^3}) \circ S_1(\frac{y^1}{y^3})(v) = e_1 + e_2 + e_3.$

Case 6.  \ \  $J = 0, y^1 = 0, y^2 = 0, y^3 \neq 0$: \ \  $\lambda(y^3)(v) = e_3.$

Case 7.  \ \  $J = 0, y^1 = 0, y^2 \neq 0, y^3 = 0$: \ \  $\lambda(y^2)(v) = e_2.$

Case 8. \ \  $J = 0, y^1 = 0, y^2 \neq 0, y^3 \neq 0$:   \ \  $\lambda(y^3) \circ S_2(\frac{y^2}{y^3})(v) = e_2 + e_3.$

\subsubsection{L(4,2,$x$,$x$), $x \neq 1$}

$$
\begin{tabular}{|c|c|c|c|c|} \hline
{\em Subalgebra } & {\em Complement} & {\em Complement}  & {$\kappa$}      & {\em Quadratic}   \\
{\lieh}        & {\compm}      & {\em Type}         & {}           & {\em Forms}        \\ \hline
$e_1$             &  $e_2 + c_1\lieh, e_3 + c_2\lieh, e_4 + c_3\lieh$      &       & 3 &  $Q$                       \\ \hline
$e_2$             &  $e_1 + c_1\lieh, e_3 + c_2\lieh, e_4 + c_3\lieh$      &       & 3 &  $Q$                       \\ \hline
$e_4$             &  $e_3, e_2, e_1$                                       & S, I  & 0 &  $Z$                       \\ \hline
$e_1 + e_2$       &  $e_4 + c_1\lieh, e_3 + c_2\lieh, e_1 + xe_2$          & R     & 2 &  $D_1, D_2, Q_{12}$        \\ \hline
\end{tabular}
$$

$$
A_1=\left( \begin{array}{c} y^1+ty^4\\ y^2\\ y^3\\ y^4 \end{array} \right), \;\;
A_2=\left( \begin{array}{c} y^1 \\ y^2+txy^4\\ y^3\\ y^4 \end{array} \right), \;\;
A_3=\left( \begin{array}{c} y^1\\ y^2\\ y^3+txy^4\\ y^4 \end{array} \right),
$$

$$
O_1(t)(v)=\left( \begin{array}{c} y^1\\ \cos(t)y^2+\sin(t)y^3\\ -\sin(t)y^2+\cos(t)y^3\\ y^4 \end{array} \right), \;\;
S_1(t)(v)=\left( \begin{array}{c} y^1/t\\ y^2\\ y^3\\ y^4 \end{array} \right).
$$

The invariant of the Adjoint action is $J=y^4$.

Case 1. \ \  $J \neq 0$: \ \ $\lambda(y^4) \circ A_3(-\frac{y^3}{xy^4}) \circ A_2(-\frac{y^2}{xy^4}) \circ A_1(-\frac{y^1}{y^4})(v) = e_4.$

Case 2. \ \  $J = 0, (y^2)^2+(y^3)^2 \neq 0, y^1 \neq 0:$
$$\lambda\left(\frac{\sqrt{(y^2)^2+(y^3)^2}}{{\rm sign}(y^2)}\right) \circ S_1\left(\frac{{\rm sign}(y^2)y^1}{\sqrt{(y^2)^2+(y^3)^2}}\right) \circ O_1\left(\arctan\frac{y^3}{y^2}\right)(v) = e_1 + e_2.$$

Case 3. \ \  $J = 0, (y^2)^2+(y^3)^2 \neq 0, y^1 = 0$:
$$\lambda\left(\frac{\sqrt{(y^2)^2+(y^3)^2}}{{\rm sign}(y^2)}\right) \circ O_1\left(\arctan\frac{y^3}{y^2}\right)(v) = e_2.$$

Case 4. \ \  $J = 0, (y^2)^2+(y^3)^2 = 0$: \ \  $\lambda(y^1)(v) = e_1.$

\subsubsection{L(4,2,1,$y$), $y \neq 1$}

$$
\begin{tabular}{|c|c|c|c|c|} \hline
{\em Subalgebra } & {\em Complement} & {\em Complement}  & {$\kappa$}      & {\em Quadratic}  \\
{\lieh}        & {\compm}      & {\em Type}         & {}           & {\em Forms}         \\ \hline
$e_1$              &  $e_2 + c_1\lieh, e_3 + c_2\lieh, e_4 + c_3\lieh$     &       & 3 &  $Q$                      \\ \hline
$e_3$              &  $e_1 + c_1\lieh, e_2 + c_2\lieh, e_4 + c_3\lieh$     &       & 3 &  $Q$                       \\ \hline
$e_4 (y \neq -1)$  &  $e_1, e_2, e_3$                                      & S, I  & 0 &  $Z$                       \\ \hline
$e_4 (y = -1)$     &  $e_2, e_1 + e_3, -e_1 + e_3$                         & S, I  & 0 &  $Q_{12}+Q_{13}, B$        \\ \hline
$e_1 + e_3$        &  $e_4 + c_1\lieh, e_2 + c_2\lieh, e_1 + ye_3$         & R     & 2 &  $D_1, D_2, Q_{12}$        \\ \hline
\end{tabular}
$$

$$
A_1(t)(v)=\left( \begin{array}{c} y^1+ty^4\\ y^2\\ y^3\\ y^4 \end{array} \right), \;\;
A_2(t)(v)=\left( \begin{array}{c} y^1\\ y^2+ty^4\\ y^3\\ y^4 \end{array} \right), \;\;
A_3(t)(v)=\left( \begin{array}{c} y^1\\ y^2\\ y^3+tyy^4\\ y^4 \end{array} \right),
$$

$$
O_1(t)(v)=\left( \begin{array}{c} \cos(t)y^1+\sin(t)y^2\\ -\sin(t)y^1+\cos(t)y^2\\ y^3\\ y^4 \end{array} \right), \;\;
S_1(t)(v)=\left( \begin{array}{c} y^1/t\\ y^2\\ y^3\\ y^4 \end{array} \right).
$$
 
The invariant of the Adjoint action is $J=y^4$.

Case 1. \ \  $J \neq 0$:  \ \   $\lambda(y^4) \circ A_3(-\frac{y^3}{yy^4}) \circ A_2(-\frac{y^2}{y^4}) \circ A_1(-\frac{y^1}{y^4})(v) = e_4.$

Case 2. \ \  $J = 0, (y^1)^2+(y^2)^2 \neq 0, y^3 \neq 0$:
 $$\lambda(y^3) \circ S_1\left(\frac{\sqrt{(y^1)^2+(y^2)^2}}{{\rm sign}(y^1)y^3}\right) \circ O_1\left(\arctan\frac{y^2}{y^1}\right)(v) = e_1 + e_3.$$

Case 3. \ \  $J = 0, (y^1)^2+(y^2)^2 \neq 0, y^3 = 0$:
$$\lambda\left(\frac{\sqrt{(y^1)^2+(y^2)^2}}{{\rm sign}(y^1)}\right) \circ O_1\left(\arctan\frac{y^2}{y^1}\right)(v) = e_1.$$

Case 4. \ \  $J = 0, (y^1)^2+(y^2)^2 = 0$:  \ \  $\lambda(y^3)(v) = e_3.$

\subsubsection{L(4,2,1,1)}

$$
\begin{tabular}{|c|c|c|c|c|} \hline
{\em Subalgebra } & {\em Complement} & {\em Complement}  & {$\kappa$}      & {\em Quadratic}  \\
{\lieh}        & {\compm}      & {\em Type}         & {}           & {\em Forms}         \\ \hline
$e_1$              &  $e_2 + c_1\lieh, e_3 + c_2\lieh, e_4 + c_3\lieh$     &       & 3 &  $Q$                     \\ \hline
$e_4$              &  $e_1, e_2, e_3$                                      & S, I  & 0 &  $Z$                         \\ \hline
\end{tabular}
$$

$$
A_1(t)(v)=\left( \begin{array}{c} y^1+ty^4\\ y^2\\ y^3\\ y^4 \end{array} \right), \;\;
A_2(t)(v)=\left( \begin{array}{c} y^1 \\ y^2+ty^4\\ y^3\\ y^4 \end{array} \right), \;\;
A_3(t)(v)=\left( \begin{array}{c} y^1\\ y^2\\ y^3+ty^4\\ y^4 \end{array} \right),
$$

$$
O_1(t)(v)=\left( \begin{array}{c} \cos(t)y^1+\sin(t)y^2\\ -\sin(t)y^1+\cos(t)y^2\\ y^3\\ y^4 \end{array} \right), \;\;
O_2(t)(v)=\left( \begin{array}{c} y^1\\ \cos(t)y^2+\sin(t)y^3\\ -\sin(t)y^2+\cos(t)y^3\\ y^4 \end{array} \right).
$$

The invariant of the Adjoint action is $J=y^4$.

Case 1. \ \  $J \neq 0:$ \ \  $\lambda(y^4) \circ A_3(-\frac{y^3}{y^4}) \circ A_2(-\frac{y^2}{y^4}) \circ A_1(-\frac{y^1}{y^4})(v) = e_4.$

Case 2. \ \  $J = 0$ \ \ $\Big(I = (y^1)^2+(y^2)^2+(y^3)^2\Big)$:
$$\lambda\left(\frac{\sqrt{I}}{{\rm sign}(y^1)}\right) \circ O_1\left(\arctan\left(\frac{\sqrt{(y^2)^2+(y^3)^2}}{{\rm sign}(y^2)y^1}\right)\right) \circ O_2\left(\arctan\frac{y^3}{y^2}\right)(v) = e_1.$$

\subsubsection{L(4,3)}

$$
\begin{tabular}{|c|c|c|c|c|} \hline
{\em Subalgebra } & {\em Complement} & {\em Complement}  & {$\kappa$}      & {\em Quadratic}    \\
{\lieh}        & {\compm}      & {\em Type}         & {}           & {\em Forms}         \\ \hline
$e_1$              &  $e_2 + c_1\lieh, e_3 + c_2\lieh, e_4 + c_3\lieh$     & R     & 3 &  $Q$                      \\ \hline
$e_2$              &  $e_4 + c\lieh, e_3, e_1$                             & I     & 2 &  $D_1, D_2, Q_{12}$       \\ \hline
$e_3$              &  $e_1 + c_1\lieh, e_2 + c_2\lieh, e_4 + c_3\lieh$     &       & 3 &  $Q$                       \\ \hline
$e_4$              &  $e_1, e_2, e_3$                                      & S, I  & 1 &  $D_2$                     \\ \hline
$e_1 + e_3$        &  $e_2 + c_1\lieh, e_4 + c_2\lieh, e_3$                & R     & 2 &  $D_1, D_2, Q_{12}$        \\ \hline
$e_2 + e_3$        &  $e_4 + c\lieh, e_3, e_1 + e_3$                       & I     & 2 &  $D_1, D_2, Q_{12}$        \\ \hline
\end{tabular}
$$

$$
A_2(t)(v)=\left( \begin{array}{c} y^1+ty^4\\ y^2\\ y^3\\ y^4 \end{array} \right), \;\;
A_3(t)(v)=\left( \begin{array}{c} y^1 \\ y^2\\ y^3+ty^4\\ y^4 \end{array} \right), \;\;
A_4(t)(v)=\left( \begin{array}{c} y^1-ty^2\\ y^2\\ e^{-t}y^3\\ y^4 \end{array} \right),
$$

$$
[O_1(t)(v)]^t=\left( \begin{array}{cccc} y^1& y^2+ty^4& y^3 & y^4 \end{array} \right), \;\;
[S_3(t)(v)]^t=\left( \begin{array}{cccc} y^1& y^2 & \frac{y^3}{t} & y^4 \end{array} \right).
$$
 
The invariants of the Adjoint action are $I = y^2$ and $J = y^4$.

Case 1. \ \  $J \neq 0$: \ \  $\lambda(y^4) \circ O_1(-\frac{y^2}{y^4}) \circ A_3(-\frac{y^3}{y^4}) \circ A_2(-\frac{y^1}{y^4})(v) = e_4.$

Case 2. \ \  $I \neq 0, J = 0, y^3 \neq 0$:  \ \  $\lambda(y^2) \circ S_3\left(\frac{y^3}{y^2}e^{-\frac{y^1}{y^2}}\right) \circ A_4(\frac{y^1}{y^2})(v) = e_2 + e_3.$

Case 3.  \ \   $I \neq 0, J = 0, y^3 = 0$:  \ \   $\lambda(y^2) \circ A_4(\frac{y^1}{y^2})(v) = e_2.$

Case 4.  \ \    $I = 0, J = 0, y^1 = 0, y^3 \neq 0$:  \ \   $\lambda(y^3)(v) = e_3.$

Case 5. \ \  $I = 0, J = 0, y^1 \neq 0, y^3 = 0:$   \  \    $\lambda(y^1)(v) = e_1.$

Case 6. \ \  $I = 0, J = 0, y^1 \neq 0, y^3 \neq 0$:  \ \   $\lambda(y^1) \circ S_3(\frac{y^3}{y^1})(v) = e_1 + e_3.$

\subsubsection{L(4,4,$x$), $x \neq 1$}

$$
\begin{tabular}{|c|c|c|c|c|} \hline
{\em Subalgebra } & {\em Complement} & {\em Complement}  & {$\kappa$}      & {\em Quadratic}   \\
{\lieh}        & {\compm}      & {\em Type}         & {}           & {\em Forms}        \\ \hline
$e_1$              &  $e_2 + c_1\lieh, e_3 + c_2\lieh, e_4 + c_3\lieh$                &       & 3 &  $Q$                       \\ \hline
$e_2$              &  $e_4 + c_1\lieh, e_3 + c_2\lieh, e_1 + e_2$                       & R     & 2 &  $D_1, D_2, Q_{12}$        \\ \hline
$e_3$              &  $e_1 + c_1\lieh, e_2 + c_2\lieh, e_4 + c_3\lieh$                &       & 3 &  $Q$                       \\ \hline
$e_4 (x \neq -1)$  &  $e_1, e_2, e_3$                                                  & S, I  & 0 &  $Z$                       \\ \hline
$e_4 (x = -1)$     &  $2e_1, -e_1 + e_2 + e_3, -e_2 + e_3$                             & S, I  & 0 &  $B$                       \\ \hline
$e_1 + e_3$        &  $e_4 + c_1\lieh, e_2 + c_2\lieh, e_1 + xe_3$                 & R     & 2 &  $D_1, D_2, Q_{12}$       \\ \hline
$e_2 + e_3$        &  $e_4 + c_1\lieh, e_1 + c_2\lieh,e_1 + e_2 + xe_3$       & R     & 2 &  $D_1, D_2, Q_{12}$        \\ \hline
\end{tabular}
$$

$$
A_1(t)(v)=\left( \begin{array}{c} y^1+ty^4\\ y^2\\ y^3\\ y^4 \end{array} \right), \;\;
A_2(t)(v)=\left( \begin{array}{c} y^1+ty^4\\ y^2+ty^4\\ y^3\\ y^4 \end{array} \right), \;\;
A_3(t)(v)=\left( \begin{array}{c} y^1\\ y^2\\ y^3+txy^4\\ y^4 \end{array} \right), \;\;
$$

$$
A_4(t)(v)=\left( \begin{array}{c} e^{-t}(y^1-ty^2)\\ e^{-t}y^2\\ e^{-tx}y^3\\ y^4 \end{array} \right), \; \;
S_3(t)(v)=\left( \begin{array}{c} y^1\\ y^2\\ y^3/t\\ y^4 \end{array} \right).
$$

The invariant of the Adjoint action is $J = y^4$.

Case 1. \ \  $J \neq 0$:  \ \
$\lambda(y^4) \circ A_3(-\frac{y^3}{xy^4}) \circ A_2(-\frac{y^2}{y^4}) \circ A_1(\frac{y^2-y^1}{y^4})(v) = e_4.$

Case 2. \  \  $J = 0, y^2 \neq 0, y^3 \neq 0$:
$\lambda\left(y^2e^{-\frac{y^1}{y^2}}\right) \circ S_3\left(\frac{y^3}{y^2}e^{-\frac{y^1}{y^2}(x-1)}\right) \circ A_4(\frac{y^1}{y^2})(v) = e_2 + e_3.$

Case 3. \ \  $J = 0, y^2 \neq 0, y^3 = 0$:  \ \   $\lambda\left(y^2e^{-\frac{y^1}{y^2}}\right) \circ A_4(\frac{y^1}{y^2})(v) = e_2.$

Case 4. \ \  $J = 0, y^2 = 0, y^1 = 0, y^3 \neq 0$:  \ \  $\lambda(y^3)(v) = e_3$.

Case 5. \  \  $J = 0, y^2 = 0, y^1 \neq 0, y^3 = 0$:  \ \   $\lambda(y^1)(v) = e_1$.

Case 6.  \  \  $J = 0, y^2 = 0, y^1 \neq 0, y^3 \neq 0$:  \ \  $\lambda(y^1) \circ S_3(\frac{y^3}{y^1})(v) = e_1 + e_3$.

\subsubsection{L(4,4,1)}

$$
\begin{tabular}{|c|c|c|c|c|} \hline
{\em Subalgebra } & {\em Complement} & {\em Complement}  & {$\kappa$}      & {\em Quadratic}    \\
{\lieh}        & {\compm}      & {\em Type}         & {}           & {\em Forms}          \\ \hline
$e_1$             &  $e_2 + c_1\lieh, e_3 + c_2\lieh, e_4 + c_3\lieh$     &       & 3 &  $Q$                       \\ \hline
$e_2$             &  $e_4 + c_1\lieh, e_3 + c_2\lieh, e_1 + e_2$          & R     & 2 &  $D_1, D_2, Q_{12}$        \\ \hline
$e_3$             &  $e_1 + c_1\lieh, e_2 + c_2\lieh, e_4 + c_3\lieh$     &       & 3 &  $Q$                       \\ \hline
$e_4$             &  $e_1, e_2, e_3$                                      & S, I  & 0 &  $Z$                       \\ \hline
\end{tabular}
$$

$$
A_1=\left( \begin{array}{c} y^1+ty^4\\ y^2\\ y^3\\ y^4 \end{array} \right), \;\;
A_2=\left( \begin{array}{c} y^1+ty^4\\ y^2+ty^4\\ y^3\\ y^4 \end{array} \right), \;\;
A_3=\left( \begin{array}{c} y^1\\ y^2\\ y^3+ty^4\\ y^4 \end{array} \right),
$$

$$
A_4=\left( \begin{array}{c} e^{-t}(y^1-ty^2)\\ e^{-t}y^2\\ e^{-t}y^3\\ y^4 \end{array} \right), \;\;
O_1=\left( \begin{array}{c} y^1+ty^3\\ y^2\\ y^3\\ y^4 \end{array} \right), \;\;
O_2=\left( \begin{array}{c} y^1\\ y^2\\ y^3+ty^2\\ y^4 \end{array} \right).
$$

The invariant of the Adjoint action is $J = y^4$.

Case 1. \  \   $J \neq 0$:  \ \  $\lambda(y^4) \circ A_3(-\frac{y^3}{y^4}) \circ A_2(-\frac{y^2}{y^4}) \circ A_1(\frac{y^2-y^1}{y^4})(v) = e_4.$

Case 2. \ \   $J = 0, y^2 \neq 0$:  \ \   $\lambda\left(y^2e^{-\frac{y^1}{y^2}}\right) \circ O_2(-\frac{y^3}{y^2}) \circ A_4(\frac{y^1}{y^2})(v) = e_2.$

Case 3. \ \  $J = 0, y^2 = 0, y^3 = 0$: \ \  $\lambda(y^1)(v) = e_1.$

Case 4. \ \  $J = 0, y^2 = 0, y^3 \neq 0$: \ \  $\lambda(y^3) \circ O_1(-\frac{y^1}{y^3})(v) = e_3.$

\subsubsection{L(4,5,$x$,$y$)}

$$
\begin{tabular}{|c|c|c|c|c|} \hline
{\em Subalgebra } & {\em Complement} & {\em Complement}  & {$\kappa$}      & {\em Quadratic}   \\
{\lieh}        & {\compm}      & {\em Type}         & {}           & {\em Forms}         \\ \hline
$e_1$              &  $e_2 + c_1\lieh, e_3 + c_2\lieh, e_4 + c_3\lieh$                 &       & 3 &  $Q$                       \\ \hline
$e_2$              &  $e_4 + c_1\lieh, e_1 + c_2\lieh, ye_2 - e_3$                     & R     & 2 &  $D_1, D_2, Q_{12}$       \\ \hline
$e_4$              &  $e_1, e_2, e_3$                                                  & S, I  & 0 &  $Z$                       \\ \hline
$e_1 + e_2$        &  $e_4 + c_1\lieh, e_1 + c_2\lieh,$                                & R     & 2 &  $D_1, D_2, Q_{12}$   \\
                   &  $xe_1 + ye_2 - e_3$                                              &       &   &                         \\ \hline
\end{tabular}
$$

$$
A_1(t)(v)=\left( \begin{array}{c} y^1+txy^4\\ y^2\\ y^3\\ y^4 \end{array} \right), \;\;
A_2(t)(v)=\left( \begin{array}{c} y^1\\ y^2+tyy^4\\ y^3-ty^4\\ y^4 \end{array} \right), \;\;
A_3(t)(v)=\left( \begin{array}{c} y^1\\ y^2+ty^4\\ y^3+tyy^4\\ y^4 \end{array} \right), \;\;
$$

$$
A_4(t)(v)=\left( \begin{array}{c} e^{-tx}y^1\\ e^{-ty}(\cos(t)y^2-\sin(t)y^3)\\ e^{-ty}(\sin(t)y^2+\cos(t)y^3)\\ y^4 \end{array} \right), \;\;
S_1(t)(v)=\left( \begin{array}{c} y^1/t\\ y^2\\ y^3\\ y^4 \end{array} \right).
$$

The invariant of the Adjoint action is $J = y^4$.

Case 1.  \ \ $J \neq 0$:  \ \   $\lambda(y^4) \circ A_3\left(-\frac{y^2+yy^3}{((y)^2+1)y^4}\right) \circ A_2\left(\frac{y^3-yy^2}{((y)^2+1)y^4}\right) \circ A_1(-\frac{y^1}{xy^4})(v) = e_4.$

Case 2.   \ \  $J = 0, \, I=(y^2)^2+(y^3)^2 \neq 0, \, y^1 \neq 0$:

$$\lambda\left(k\right) \circ S_1\left(\frac{{\rm sign}(y^2)y^1}{\sqrt{I}}e^{\left(\arctan\frac{y^3}{y^2}\right)(x-y)}\right)\circ A_4\left(-\arctan\frac{y^3}{y^2}\right)(v)  = e_1+e_2.$$

\hspace{2cm} Note: $\displaystyle k=\frac{\sqrt{I}}{{\rm sign}(y^2)}\exp\left[{\left(\arctan\frac{y^3}{y^2}\right)y}\right].$

\

Case 3. \ \   $J = 0, \, (y^2)^2+(y^3)^2 \neq 0,\,  y^1 = 0:$  \ \   $\lambda(y^3e^{\frac{\pi}{2}y}) \circ A_4(-\frac{\pi}{2})(v) = e_2.$

Case 4. \ \   $J = 0, \, (y^2)^2+(y^3)^2 = 0:$ \ \   $\lambda(y^1)(v) = e_1.$

\subsubsection{L(4,6)}

$$
\begin{tabular}{|c|c|c|c|c|} \hline
{\em Subalgebra } & {\em Complement} & {\em Complement}  & {$\kappa$}      & {\em Quadratic}    \\
{\lieh}        & {\compm}      & {\em Type}         & {}           & {\em Forms}         \\ \hline
$e_1$             &  $e_2 + c_1\lieh, e_3 + c_2\lieh, e_4 + c_3\lieh$     &       & 3 &  $Q$                      \\ \hline
$e_2$             &  $e_4 + c_1\lieh, e_3 + c_2\lieh, e_1 + e_2$          & R     & 2 &  $D_1, D_2, Q_{12}$       \\ \hline
$e_3$             &  $e_4 + c_1\lieh, e_1 + c_2\lieh, e_2 + e_3$          & R     & 2 &  $D_1, D_2, Q_{12}$       \\ \hline
$e_4$             &  $e_1, e_2, e_3$                                      & S, I  & 0 &  $Z$                      \\ \hline
\end{tabular}
$$

$$
A_1(t)(v)=\left( \begin{array}{c} y^1+ty^4\\ y^2\\ y^3\\ y^4 \end{array} \right), \;\;
A_2(t)(v)=\left( \begin{array}{c} y^1+ty^4\\ y^2+ty^4\\ y^3\\ y^4 \end{array} \right), \;\;
A_3(t)(v)=\left( \begin{array}{c} y^1\\ y^2+ty^4\\ y^3+ty^4\\ y^4 \end{array} \right), \;\;
$$

$$
A_4(t)(v)=\left( \begin{array}{c} e^{-t}(y^1-ty^2+\frac{1}{2}t^2y^3)\\ e^{-t}(y^2-ty^3)\\ e^{-t}y^3\\ y^4 \end{array} \right),
O_1(t)(v)=\left( \begin{array}{c} y^1+ty^3\\ y^2\\ y^3\\ y^4 \end{array} \right).
$$

The invariant of the Adjoint action is $J = y^4$.

Case 1. \ \  $J \neq 0$ \ \   $\lambda(y^4) \circ A_3(-\frac{y^3}{y^4}) \circ A_2(\frac{y^3-y^2}{y^4}) \circ A_1(-\frac{y^1-y^2+y^3}{y^4})(v) = e_4.$

Case 2. \ \ $J = 0, y^3 \neq 0$:  \ \  $\lambda\left(y^3e^{-\frac{y^2}{y^3}}\right) \circ O_1\left(\frac{\frac{1}{2}(y^2)^2-y^1y^3}{(y^3)^2}\right) \circ A_4(\frac{y^2}{y^3})(v) = e_3.$

Case 3. \ \  $J = 0, y^2 \neq 0, y^3 = 0$:  \ \  $\lambda\left(y^2e^{-\frac{y^1}{y^2}}\right) \circ A_4(\frac{y^1}{y^2})(v) = e_2.$

Case 4. \ \  $J = 0, y^2 = 0, y^3 = 0:$ \ \  $\lambda(y^1)(v) = e_1.$

\subsubsection{L(4,7)}

$$
\begin{tabular}{|c|c|c|c|c|c|} \hline
{\em Subalgebra } & {\em Complement} & {\em C}  & {$\kappa$}      & {\em Quadratic}   & {\em Petrov} \\
{\lieh}        & {\compm}      & {\em Type}         & {}           & {\em Forms}        & {\em Number}  \\ \hline
$e_1$             &  $e_2 + c_1\lieh, e_3 + c_2\lieh, e_4 + c_3\lieh$     &       & 3 &  $Q$                  &                 \\ \hline
$e_2$             &  $e_4 + c_1\lieh, e_3 + c_2\lieh, e_1 + c_3\lieh$     &       & 2 &  $D_1, D_2, Q_{12}$   &                 \\ \hline
$e_3$             &  $e_4 + c\lieh, e_2, -e_1$                            & I     & 2 &  $D_1, D_2, Q_{12}$   &                 \\ \hline
$e_4$             &  $e_3 + c\lieh, e_2, e_1$                             & I     & 1 &  $D_1$                &                 \\ \hline
$e_2 + e_3$       &  $e_4 + c\lieh, e_2, e_1$                             & I     & 1 &  $D_1, N$             &  32.14  \\
& & & & & ($c=1$)  \\ \hline
$e_3 + e_4$       &  $e_3 + c\lieh, e_2, e_1$                             & I     & 1 &  $D_1$                &                 \\ \hline
\end{tabular}
$$

$$
A_1=\left( \begin{array}{c} y^1+ty^4\\ y^2\\ y^3\\ y^4 \end{array} \right), \;
A_2=\left( \begin{array}{c} y^1+ty^3\\ y^2+ty^4\\ y^3\\ y^4 \end{array} \right), \;
A_3=\left( \begin{array}{c} y^1-ty^2\\ y^2\\ y^3\\ y^4 \end{array} \right), \;
S_{13}=\left( \begin{array}{c} y^1/t \\ y^2 \\ y^3/t \\ y^4 \end{array} \right).
$$

The invariants of the Adjoint action are $I = y^3, J = y^4$.

Case 1. \ \  $I \neq 0, J \neq 0$:  \ \   $\lambda(y^4) \circ S_{13}(\frac{y^3}{y^4}) \circ A_2(-\frac{y^2}{y^4}) \circ A_1(\frac{y^2y^3-y^1y^4}{(y^4)^2})(v) = e_3 + e_4.$

Case 2. \ \   $I = 0, J \neq 0$:  \ \  $\lambda(y^4) \circ A_2(-\frac{y^2}{y^4}) \circ A_1(-\frac{y^1}{y^4})(v) = e_4.$

Case 3. \ \  $I \neq 0, J = 0, y^2 \neq 0$: \ \  $\lambda(y^2) \circ S_{13}(\frac{y^3}{y^2}) \circ A_2(-\frac{y^1}{y^3})(v) = e_2 + e_3.$

Case 4. \ \  $I \neq 0, J = 0, y^2 = 0$: \ \  $\lambda(y^3) \circ A_2(-\frac{y^1}{y^3})(v) = e_3.$

Case 5. \ \  $I = 0, J = 0, y^2 \neq 0$:  \ \  $\lambda(y^2) \circ A_3(\frac{y^1}{y^2})(v) = e_2.$

Case 6. \ \  $I = 0, J = 0, y^2 = 0$: \ \  $\lambda(y^1)(v) = e_1.$

\subsubsection{L(4,8)}

$$
\begin{tabular}{|c|c|c|c|c|c|} \hline
{\em Subalgebra } & {\em Complement} & {\em C}  & {$\kappa$}      & {\em Quadratic}   & {\em Petrov} \\
{\lieh}        & {\compm}      & {\em Type}         & {}           & {\em Forms}        & {\em Number}  \\ \hline
$e_1$             &  $e_2 + c_1\lieh, e_3 + c_2\lieh, e_4 + c_3\lieh$     & R     & 3 &  $Q$                  &                 \\ \hline
$e_2$             &  $e_4 + c_1\lieh, e_3 + c_2\lieh, e_1 + c_3\lieh$     &       & 2 &  $D_1, D_2, Q_{12}$   &                 \\ \hline
$e_3$             &  $e_4 + c_1\lieh, e_2 + c_2\lieh, -e_1 - c_3\lieh$    &       & 2 &  $D_1, D_2, Q_{12}$   &                 \\ \hline
$e_4$             &  $e_1, e_2 + e_3, -e_2 + e_3 $                        & I     & 1 &  $D_1, B$             &  32.3           \\ \hline
$e_2 + e_3$       &  $e_4, e_2 - e_3, 2e_1$                               & S     & 1 &  $D_1, N$             &  32.14 \\
& & & & &  ($c=0$)  \\ \hline
\end{tabular}
$$

$$
A_2=\left( \begin{array}{c} y^1+ty^3\\ y^2+ty^4\\ y^3\\ y^4 \end{array} \right), \;
A_3=\left( \begin{array}{c} y^1-ty^2\\ y^2\\ y^3-ty^4\\ y^4 \end{array} \right), \;
O_1=\left( \begin{array}{c} y^1+ty^4\\ y^2\\ y^3\\ y^4 \end{array} \right), \;
S_{12}=\left( \begin{array}{c} {y^1}/{t} \\ {y^2}/{t} \\ y^3 \\ y^4 \end{array} \right).
$$

The invariants of the Adjoint action are $I = y^2y^3 - y^1y^4$ and $J = y^4$.

Case 1. \ \  $J \neq 0$:  \ \  $\lambda(y^4) \circ O_1\left(\frac{I}{(y^4)^2}\right) \circ A_3(\frac{y^3}{y^4}) \circ A_2(-\frac{y^2}{y^4})(v) = e_4.$

Case 2. \ \  $I \neq 0, J = 0$:  \ \ $\lambda(y^3) \circ S_{12}(\frac{y^2}{y^3}) \circ A_2(-\frac{y^1}{y^3})(v) = e_2 + e_3.$

Case 3. \ \  $I = 0, J = 0, y^2 = 0, y^3 \neq 0$:  \ \  $\lambda(y^3) \circ A_2(-\frac{y^1}{y^3})(v) = e_3.$

Case 4. \ \  $I = 0, J = 0, y^2 \neq 0, y^3 = 0$:  \ \  $\lambda(y^2) \circ A_3(\frac{y^1}{y^2})(v) = e_2.$

Case 5. \ \  $I = 0, J = 0, y^2 = 0, y^3 = 0:$  \ \  $\lambda(y^1)(v) = e_1.$

\subsubsection{L(4,9,$x$), $x \neq 1$}

$$
\begin{tabular}{|c|c|c|c|c|c|} \hline

{\em Subalgebra } & {\em Complement} & {\em C}  & {$\kappa$}      & {\em Quadratic}   & {\em Petrov} \\
{\lieh}        & {\compm}      & {\em Type}           &        & {\em Forms}        & {\em Number}  \\ \hline

$e_1$             &  $e_2 + c_1\lieh, e_3 + c_2\lieh, e_4 + c_3\lieh$     &       & 3 &  $Q$                  &                \\ \hline
$e_2$             &  $e_4 + c_1\lieh, e_3 + c_2\lieh, e_1 + c_3\lieh$     &       & 2 &  $D_1, D_2, Q_{12}$   &                \\ \hline
$e_3$             &  $e_4 + c_1\lieh, e_2 + c_2\lieh, -e_1 + c_3\lieh$    &       & 2 &  $D_1, D_2, Q_{12}$   &                \\ \hline
$e_4$             &  $e_1, e_2, e_3$                                      & I     & 0 &  $Z$                  &                \\ \hline
$e_2 + e_3$       &  $\frac{1}{1-x}(e_4+c\lieh), \frac{1}{1-x}(e_2+xe_3),$ & R & 1 &  $D_1, N$            &     32.14      \\ 
 & $e_1$ & & & &  $(c \neq 0,1)$ \\ \hline

\end{tabular}
$$

$$
A_1=\left( \begin{array}{c} y^1+t(x+1)y^4\\ y^2\\ y^3\\ y^4 \end{array} \right), \;\;
A_2=\left( \begin{array}{c} y^1+ty^3\\ y^2+ty^4\\ y^3\\ y^4 \end{array} \right), \;\;
A_3=\left( \begin{array}{c} y^1-ty^2\\ y^2\\ y^3+txy^4\\ y^4 \end{array} \right),
$$

$$
[S_{12}]^t=\left( \begin{array}{cccc} \frac{y^1}{t} & \frac{y^2}{t} & y^3 & y^4 \end{array} \right).
$$

The invariant of the Adjoint action is $J = y^4$.

Case 1. \ \  $J \neq 0$:  \ \  $\lambda(y^4) \circ A_3(-\frac{y^3}{xy^4}) \circ A_2(-\frac{y^2}{y^4}) \circ A_1\left(\frac{y^2y^3-y^1y^4}{(1+x)(y^4)^2}\right)(v) = e_4.$

Case 2. \ \  $J = 0, y^2 = 0, y^3 \neq 0$:  \ \  $\lambda(y^3) \circ A_2(-\frac{y^1}{y^3})(v) = e_3.$

Case 3. \ \  $J = 0, y^2 \neq 0, y^3 \neq 0$:  \ \  $\lambda(y^3) \circ S_{12}(\frac{y^2}{y^3}) \circ A_2(-\frac{y^1}{y^3})(v) = e_2 + e_3.$

Case 4. \ \  $J = 0, y^2 \neq 0, y^3 = 0$:  \ \  $\lambda(y^2) \circ A_3(\frac{y^1}{y^2})(v) = e_2$

Case 5. \ \  $J = 0, y^2 = 0, y^3 = 0:$ \ \  $\lambda(y^1)(v) = e_1.$

\subsubsection{L(4,9,1)}

$$
\begin{tabular}{|c|c|c|c|c|} \hline

{\em Subalgebra } & {\em Complement} & {\em Complement}  & {$\kappa$}      & {\em Quadratic}    \\
{\lieh}        & {\compm}      & {\em Type}         & {}           & {\em Forms}         \\ \hline

$e_1$             &  $e_2 + c_1\lieh, e_3 + c_2\lieh, e_4 + c_3\lieh$     &       & 3 &  $Q$                                 \\ \hline
$e_2$             &  $e_4 + c_1\lieh, e_3 + c_2\lieh, e_1 + c_3\lieh$     &       & 2 &  $D_1, D_2, Q_{12}$                  \\ \hline
$e_4$             &  $e_1, e_2, e_3$                                      & I     & 0 &  $Z$                                 \\ \hline

\end{tabular}
$$

$$
A_1(t)(v)=\left( \begin{array}{c} y^1+2ty^4\\ y^2\\ y^3\\ y^4 \end{array} \right), \;\;
A_2(t)(v)=\left( \begin{array}{c} y^1+ty^3\\ y^2+ty^4\\ y^3\\ y^4 \end{array} \right), \;\;
A_3(t)(v)=\left( \begin{array}{c} y^1-ty^2\\ y^2\\ y^3+ty^4\\ y^4 \end{array} \right), \;\;
$$

$$
[O_1(t)(v)]^t=\left( \begin{array}{cccc} y^1 &  \cos(t)y^2+\sin(t)y^3  &  -\sin(t)y^2+\cos(t)y^3 & y^4 \end{array} \right).
$$

The invariant of the Adjoint action is $J = y^4$.

Case 1. \ \  $J \neq 0$: \ \   $\lambda(y^4) \circ A_3(-\frac{y^3}{y^4}) \circ A_2(-\frac{y^2}{y^4}) \circ A_1\left(\frac{y^2y^3-y^1y^4}{2(y^4)^2}\right)(v) = e_4$.

Case 2.  \ \  $J = 0, \,\, I = (y^2)^2+(y^3)^2 \neq 0$: \ \ 

\begin{center}
$\lambda\left(\frac{\sqrt{I}}{{\rm sign}(y^2)}\right) \circ O_1\left(\arctan\frac{y^3}{y^2}\right) \circ A_3\left(\frac{y^1y^2}{I}\right) \circ A_2\left(-\frac{y^1y^3}{I}\right)(v) = e_2.$
\end{center}

Case 3.  \ \ $J = 0, \,  (y^2)^2+(y^3)^2 = 0$:  \ \  $\lambda(y^1)(v) = e_1$. \\

\subsubsection{L(4,10)}

$$
\begin{tabular}{|c|c|c|c|c|c|} \hline

{\em Subalgebra } & {\em Complement} & {\em C}  & {$\kappa$}      & {\em Quadratic}   & {\em Petrov} \\
{\lieh}        & {\compm}      & {\em Type}         & {}           & {\em Forms}        & {\em Number}  \\ \hline

$e_1$             &  $e_2 + c_1\lieh, e_3 + c_2\lieh, e_4 + c_3\lieh$     &     & 3 &  $Q$                  &                \\ \hline
$e_2$             &  $e_4 + c_1\lieh, e_3 + c_2\lieh, e_1 + c_3\lieh$     &     & 2 &  $D_1, D_2, Q_{12}$   &                \\ \hline
$e_3$             &  $e_4 + c\lieh, e_2 + e_3, e_1$                       &  R  & 1 &  $D_1, N$             &   32.15        \\ \hline
$e_4$             &  $e_1, e_2, e_3$                                      &  I  & 0 &  $Z$                  &                \\ \hline

\end{tabular}
$$

$$
A_2=\left( \begin{array}{c} y^1+ty^3\\ y^2+ty^4\\ y^3\\ y^4 \end{array} \right),\;\;
A_3=\left( \begin{array}{c} y^1-ty^2-\frac{1}{2}t^2y^4\\ y^2+ty^4\\ y^3+ty^4\\ y^4 \end{array} \right), \; \; 
A_4=\left( \begin{array}{c} e^{-2t}y^1\\ e^{-t}(y^2-ty^3)\\ e^{-t}y^3\\ y^4 \end{array} \right).
$$

The invariant of the Adjoint action is $J = y^4$.

Case 1. \ \   $J \neq 0$:  \ \ \Big( $r = 2y^1y^4-2y^2y^3+3(y^3)^2$\Big)

\begin{center}
$\lambda(y^4) \circ A_3(-\frac{1+y^3}{y^4}) \circ A_2(-\frac{1}{4}\frac{r}{y^4}) \circ A_3(\frac{1}{y^4}) \circ A_2\left(\frac{1}{4}\frac{r}{y^4}-\frac{y^2-y^3}{y^4}\right)(v) = e_4.$
\end{center}

Case 2. \ \  $J = 0, y^3 \neq 0$:  \ \   $\lambda\left(y^3e^{-\frac{y^2}{y^3}}\right) \circ A_4(\frac{y^2}{y^3}) \circ A_2(-\frac{y^1}{y^3})(v) = e_3.$

Case 3. \ \  $J = 0, y^3 = 0, y^2 \neq 0$:  \ \ $\lambda(y^2) \circ A_3(\frac{y^1}{y^2})(v) = e_2$. 

Case 4. \ \  $J = 0, y^3 = 0, y^2 = 0$: \ \  $\lambda(y^1)(v) = e_1$.

\subsubsection{L(4,11)}

$$
\begin{tabular}{|c|c|c|c|c|c|} \hline

{\em Subalgebra } & {\em Complement} & {\em C}  & {$\kappa$}      & {\em Quadratic}   & {\em Petrov} \\
{\lieh}        & {\compm}      & {\em Type}         & {}           & {\em Forms}        & {\em Number}  \\ \hline

$e_1$             &  $e_2 + c_1\lieh, e_3 + c_2\lieh, e_4 + c_3\lieh$     &  R  & 3 &  $Q$                  &                 \\ \hline
$e_2$             &  $-e_4, e_3, -e_1$                                    &  S  & 1 &  $D_1, N$             &  32.16 ($q=0$)  \\ \hline
$e_4$             &  $e_1, e_2, -e_3$                                     &  I  & 1 &  $D_1, R$             &  32.4           \\ \hline

\end{tabular}
$$

$$
A_2(t)(v)=\left( \begin{array}{c} y^1+ty^3-\frac{1}{2}t^2y^4\\ y^2\\ y^3-ty^4\\ y^4 \end{array} \right), \;\;
A_3(t)(v)=\left( \begin{array}{c} y^1-ty^2-\frac{1}{2}t^2y^4\\ y^2+ty^4\\ y^3\\ y^4 \end{array} \right), \;\;
$$

$$
A_4(t)(v)=\left( \begin{array}{c} y^1\\ \cos(t)y^2-\sin(t)y^3\\ \sin(t)y^2+\cos(t)y^3\\ y^4 \end{array} \right), \;\;
O_1(t)(v)=\left( \begin{array}{c} y^1+ty^4\\ y^2\\ y^3\\ y^4 \end{array} \right).
$$

The invariants of the Adjoint action are $I = 2y^1y^4 + (y^2)^2 + (y^3)^2, \, J = y^4$.

Case 1. \ \  $J \neq 0$: \ \  $\lambda(y^4) \circ O_1\left(-\frac{1}{2}\frac{I}{(y^4)^2}\right) \circ A_3(-\frac{y^2}{y^4}) \circ A_2(\frac{y^3}{y^4})(v) = e_4.$

Case 2. \ \  $J = 0, I \neq 0$: \ \  

 \hspace{1cm} $\lambda\left(\frac{\sqrt{I}}{{\rm sign}(y^2)}\right) \circ A_4\left(-\arctan\frac{y^3}{y^2}\right) \circ A_3(\frac{y^1y^2}{I}) \circ A_2(-\frac{y^1y^3}{I})(v) = e_2.$

Case 3. \ \  $J = 0, I = 0$:  \ \  $\lambda(y^1)(v) = e_1.$

\subsubsection{L(4,12,$x$)}

$$
\begin{tabular}{|c|c|c|c|c|c|} \hline

{\em Subalgebra } & {\em Complement} & {\em C}  & {$\kappa$}      & {\em Quadratic}   & {\em Petrov} \\
{\lieh}        & {\compm}      & {\em Type}         & {}           & {\em Forms}        & {\em Number}  \\ \hline

$e_1$             &  $e_2 + c_1\lieh, e_3 + c_2\lieh, e_4 + c_3\lieh$    &     & 3 &  $Q$                  &                     \\ \hline
$e_2$             &  $e_4 + c\lieh, xe_2-e_3, e_1$                       &  R  & 1 &  $D_1, N$             &  32.16 ($q \neq 0$) \\ \hline
$e_4$             &  $e_1, e_2, e_3$                                     &  I  & 0 &  $Z$                  &                     \\ \hline

\end{tabular}
$$

$$
A_1(t)(v)=\left( \begin{array}{c} y^1+2txy^4\\ y^2\\ y^3\\ y^4 \end{array} \right), \;\;
A_2(t)(v)=\left( \begin{array}{c} y^1+ty^3-\frac{1}{2}t^2y^4\\ y^2+txy^4\\ y^3-ty^4\\ y^4 \end{array} \right), \;\;
$$

$$
A_3(t)(v)=\left( \begin{array}{c} y^1-ty^2-\frac{1}{2}t^2y^4\\ y^2+ty^4\\ y^3+txy^4\\ y^4 \end{array} \right), \;\;
A_4(t)(v)=\left( \begin{array}{c} e^{-2tx}y^1\\ e^{-tx}(\cos(t)y^2-\sin(t)y^3)\\ e^{-tx}(\sin(t)y^2+\cos(t)y^3)\\ y^4 \end{array} \right).
$$
 
The invariant of the Adjoint action is $J = y^4$.

Case 1. \ \  $J \neq 0$:  \ \ $\left(r = -\frac{1}{4}\frac{2y^1y^4(x^2+1)^2 + (y^3)^2(3x^2+1) + ((y^2)^2+2y^2y^3x)(1-x^2)}{x(x^2+1)^2(y^4)^2} \right)$

\begin{center}
$\lambda(y^4) \circ A_3\left(-\frac{y^2+xy^3}{(x^2+1)y^4}\right) \circ A_2\left(\frac{y^3-xy^2}{(x^2+1)y^4}\right) \circ A_1(r)(v) = e_4$.
\end{center}

Case 2. \ \  $J = 0, \, I = (y^2)^2+(y^3)^2 \neq 0$:  \ \

\begin{center}
$\lambda\left(\frac{\sqrt{I}}{{\rm sign}(y^2)}e^{\left(\arctan\frac{y^3}{y^2}\right) x}\right) \circ A_4\left(-\arctan\frac{y^3}{y^2}\right) \circ A_3\left(\frac{y^1y^2}{I}\right) \circ A_2\left(-\frac{y^1y^3}{I}\right)(v) = e_2$.
\end{center}

Case 3.  \  \  $J = 0, (y^2)^2+(y^3)^2 = 0$:  \ \  $\lambda(y^1)(v) = e_1$.

\subsubsection{L(4,13)}

$$
\begin{tabular}{|c|c|c|c|c|c|} \hline

{\em Subalgebra } & {\em Complement} & {\em C}  & {$\kappa$}      & {\em Quadratic}   & {\em Petrov} \\
{\lieh}        & {\compm}      & {\em Type}         & {}           & {\em Forms}        & {\em Number}  \\ \hline

$e_1$             &  $e_4 + c_1\lieh, e_3 + c_2\lieh, -e_2 + c_3\lieh$   &     & 2 &  $D_1, D_2, Q_{12}$   &                     \\ \hline
$e_3+ke_4$        &  $e_4 + c\lieh, e_2, e_1$                            &  I  & 1 &  $D_1$                &                     \\ \hline
$e_4$             &  $e_3 + c\lieh, e_2, e_1$                            &  I  & 1 &  $D_1, R$             &    32.6             \\ \hline

\end{tabular}
$$

$$
A_1=\left( \begin{array}{c} y^1+ty^3\\ y^2-ty^4\\ y^3\\ y^4 \end{array} \right),
A_2=\left( \begin{array}{c} y^1+ty^4\\ y^2+ty^3\\ y^3\\ y^4 \end{array} \right),
A_4=\left( \begin{array}{c} \cos(t)y^1-\sin(t)y^2\\ \sin(t)y^1+\cos(t)y^2\\ y^3\\ y^4 \end{array} \right).
$$

The invariants of the Adjoint action are $I = y^3, J = y^4$.

Case 1. \ \  $I \neq 0$:  \ \  $\lambda(y^3) \circ A_2\left(-\frac{y^1y^4+y^2y^3}{(y^3)^2+(y^4)^2}\right) \circ A_1\left(-\frac{y^1y^3-y^2y^4}{(y^3)^2+(y^4)^2}\right) = e_3 + \frac{y^4}{y^3}e_4.$

Case 2. \ \  $I = 0, J \neq 0$:  \ \  $\lambda(y^4) \circ A_2(-\frac{y^1}{y^4}) \circ A_1(\frac{y^2}{y^4}) = e_4.$

Case 3. \ \  $I = 0, J = 0$:  \ \  $\lambda\left(\frac{\sqrt{(y^1)^2+(y^2)^2}}{{\rm sign}(y^1)}\right) \circ A_4\left(-\arctan\frac{y^2}{y^1}\right) = e_1$

\subsubsection{L(4,-1)}

$$
\begin{tabular}{|c|c|c|c|c|} \hline

{\em Subalgebra } & {\em Complement} & {\em Complement}  & {$\kappa$}      & {\em Quadratic}  \\
{\lieh}        & {\compm}      & {\em Type}         & {}           & {\em Forms}          \\ \hline
$e_1$       & $e_2 + c_1\lieh, e_3 + c_2\lieh, e_4 + c_3\lieh$ &      & 3 & $Q$                         \\ \hline
$e_2$       & $e_4, e_3, e_1$                                  & S, I & 2 & $D_1, D_2, Q_{12}$          \\ \hline
$e_3$       & $e_1, e_2 + c_1\lieh, e_4 + c_2\lieh$            & I    & 3 & $Q$                         \\ \hline
$e_2 + e_3$ & $e_4 + c_1\lieh, e_2 + c_2\lieh, e_1$            & I    & 2 & $D_1, D_2, Q_{12}$          \\ \hline
\end{tabular}
$$

$$
A_1(t)(v)=\left( \begin{array}{c} y^1+ty^2\\ y^2\\ y^3\\ y^4 \end{array} \right), \;\;
O_1(t)(v)=\left( \begin{array}{c} y^1\\ y^2\\ y^3+ty^2\\ y^4 \end{array} \right), \;\;
O_2(t)(v)=\left( \begin{array}{c} y^1\\ y^2\\ y^3\\ y^4+ty^2 \end{array} \right),
$$

$$
O_3(t)(v)=\left( \begin{array}{c} y^1\\ y^2\\ \cos(t)y^3+\sin(t)y^4\\ -\sin(t)y^3+\cos(t)y^4 \end{array} \right), \;\;
S_1(t)(v)=\left( \begin{array}{c} y^1/t\\ y^2\\ y^3\\ y^4 \end{array} \right).
$$

The invariants of the Adjoint action are $H = y^2, I=y^3,$ and $J=y^4$.

Case 1.  \ \  $H \neq 0$:  \ \  $\lambda(y^2) \circ O_2(-\frac{y^4}{y^2}) \circ O_1(-\frac{y^3}{y^2}) \circ A_1(-\frac{y^1}{y^2})(v) = e_2.$

Case 2.  \ \ $H = 0, I^2+J^2 \neq 0, y^1 \neq 0$:

\begin{center}
$\lambda\left(\frac{\sqrt{(y^3)^2+(y^4)^2}}{{\rm sign}(y^3)}\right) \circ S_1\left(\frac{{\rm sign}(y^3)y^1}{\sqrt{(y^3)^2+(y^4)^2}}\right) \circ O_3\left(\arctan\frac{y^4}{y^3}\right)(v) = e_1 + e_3. $
\end{center}

Case 3. \ \  $H = 0, I^2+J^2 \neq 0, y^1 = 0$: \ \  $\lambda\left(\frac{\sqrt{(y^3)^2+(y^4)^2}}{{\rm sign}(y^3)}\right) \circ O_3\left(\arctan\frac{y^4}{y^3}\right)(v) = e_3.$

Case 4. \ \ $H = 0, I^2+J^2 = 0$:  \ \  $\lambda(y^1)(v) = e_1.$

\subsubsection{L(4,-2)}

$$
\begin{tabular}{|c|c|c|c|c|c|} \hline

{\em Subalgebra } & {\em Complement} & {\em C}  & {$\kappa$}      & {\em Quadratic}   & {\em Petrov} \\
{\lieh}        & {\compm}      & {\em Type}         & {}           & {\em Forms}        & {\em Number}  \\ \hline

$e_1$        & $e_2 + c_1\lieh, e_3 + c_2\lieh, e_4 + c_3\lieh$      &     & 3 & $Q$                &         \\ \hline
$e_2$        & $e_4 + c\lieh, e_3, e_1$                              &  I  & 2 & $D_1, D_2, Q_{12}$ &         \\ \hline
$e_3$        & $e_1 + c_1\lieh, e_2 + c_2\lieh, e_4 + c_3\lieh$      &     & 3 & $Q$                &         \\ \hline
$e_4$        & $e_1, e_2 + c\lieh, e_3$                              &  I  & 2 & $D_1, D_2, Q_{12}$ &         \\ \hline
$e_1 + e_3$  & $e_2 + e_4 + c_1\lieh, -e_2+c_2\lieh,$                &     & 2 & $D_1, D_2, Q_{12}$ &         \\ 
             & $e_3 + c_3\lieh$                                      &     &   &                    &         \\ \hline
$e_2 + e_3$  & $e_4 + c\lieh, e_3, e_1$                              &  I  & 1 & $D_1$              &         \\ \hline
$e_1 + e_4$  & $e_2 + c\lieh, e_1, e_3$                              &  I  & 1 & $D_1$              &         \\ \hline

$e_2 + ke_4$ & $e_4 + c_1\lieh, e_2 + c_2\lieh, e_1$                 &  I  & 1 & $D_1$              &         \\
($k \neq -1,0$) &                                                    &     &   &                    &         \\ \hline
$e_2 - e_4$  & $e_4 + c_1\lieh, e_2 + c_2\lieh, e_1$                 &  I  & 1 & $D_1, B$           &  32.5   \\ \hline

\end{tabular}
$$

$$
A_1=\left( \begin{array}{c} y^1+ty^2\\ y^2\\ y^3\\ y^4 \end{array} \right), \;\;
A_3=\left( \begin{array}{c} y^1\\ y^2\\ y^3+ty^4\\ y^4 \end{array} \right), \;\;
S_1=\left( \begin{array}{c} \frac{y^1}{t} \\ y^2 \\ y^3 \\ y^4 \end{array} \right), \;\;
S_3=\left( \begin{array}{c} y^1 \\ y^2 \\ \frac{y^3}{t} \\ y^4 \end{array} \right).
$$

The invariants of the Adjoint action are $I = y^2$ and $J = y^4$.

Case 1. \ \  $I \neq 0, J \neq 0$:  \ \ $\lambda(y^2) \circ A_3(-\frac{y^3}{y^4}) \circ A_1(-\frac{y^1}{y^2})(v) = e_2 + \frac{y^4}{y^2}e_4.$

Case 2. \ \  $I \neq 0, J = 0, y^3 \neq 0$: \ \  $\lambda(y^2) \circ S_3(\frac{y^3}{y^2}) \circ A_1(-\frac{y^1}{y^2})(v) = e_2 + e_3.$

Case 3. \ \  $I \neq 0, J = 0, y^3 = 0$:  \ \  $\lambda(y^2) \circ A_1(-\frac{y^1}{y^2})(v) = e_2.$

Case 4. \ \  $I = 0, J \neq 0, y^1 \neq 0$: \ \  $\lambda(y^4) \circ S_1(\frac{y^1}{y^4}) \circ A_3(-\frac{y^3}{y^4})(v) = e_1 + e_4.$

Case 5. \ \  $I = 0, J \neq 0, y^1 = 0$:  \ \  $\lambda(y^4) \circ A_3(-\frac{y^3}{y^4})(v) = e_4.$

Case 6. \ \  $I = 0, J = 0, y^1y^3 \neq 0$:  \ \  $\lambda(y^3) \circ S_1(\frac{y^1}{y^3})(v) = e_1 + e_3.$

Case 7. \ \  $I = 0, J = 0, y^1 \neq 0$: \ \  $\lambda(y^1)(v) = e_1.$

Case 8. \ \  $I = 0, J = 0, y^3 \neq 0$:  \ \  $\lambda(y^3)(v) = e_3.$

\subsubsection{L(4,-3)}

$$
\begin{tabular}{|c|c|c|c|c|} \hline

{\em Subalgebra } & {\em Complement} & {\em Complement}  & {$\kappa$}      & {\em Quadratic}   \\
{\lieh}        & {\compm}      & {\em Type}         & {}           & {\em Forms}          \\ \hline

$e_1$       & $e_2 + c_1\lieh, e_3 + c_2\lieh, e_4 + c_3\lieh$ & S    & 3 & $Q$                         \\ \hline
$e_2$       & $e_4, e_3 + c\lieh, e_1$                         & S, I & 2 & $D_1, D_2, Q_{12}$          \\ \hline
$e_4$       & $e_1, e_2 + c_1\lieh, e_3 + c_2\lieh$            & I    & 3 & $Q$                       \\ \hline

\end{tabular}
$$

$$
A_2(t)(v)=\left( \begin{array}{c} y^1+ty^3\\ y^2\\ y^3\\ y^4 \end{array} \right), \;
A_3(t)(v)=\left( \begin{array}{c} y^1-ty^2\\ y^2\\ y^3\\ y^4 \end{array} \right), \;
O_1(t)(v)=\left( \begin{array}{c} y^1+ty^4\\ y^2\\ y^3\\ y^4 \end{array} \right),
$$

$$
O_2(t)(v)=\left( \begin{array}{c} y^1 \\ \cos(t)y^2+\sin(t)y^3 \\ -\sin(t)y^2+\cos(t)y^3 \\ y^4 \end{array} \right), \;\;
O_3(t)(v)=\left( \begin{array}{c} y^1\\ y^2\\ y^3\\ y^4+ty^2 \end{array} \right).
$$

The invariants of the Adjoint action are $H=y^2, I=y^3,$ and $J=y^4$.

Case 1. \ \  $H^2+I^2 \neq 0$: \ \

\begin{center}
$\lambda(k) \circ O_3\left(-\frac{{\rm sign}(y^2)y^4}{\sqrt{H^2+I^2}}\right) \circ O_2\left(\arctan\frac{y^3}{y^2}\right) \circ A_3\left(\frac{y^1y^2}{H^2+I^2}\right) \circ A_2\left(-\frac{y^1y^3}{H^2+I^2}\right)(v) = e_2.$
\end{center}

\hspace{2cm} Note: $k=\frac{\sqrt{H^2+I^2}}{{\rm sign}(y^2)}$.

Case 2.  \ \  $H^2+I^2 = 0, J \neq 0$:  \ \  $\lambda(y^4) \circ O_1(-\frac{y^1}{y^4})(v) = e_4$.

Case 3. \ \  $H^2+I^2 = 0, J = 0$:  \ \  $\lambda(y^1)(v) = e_1$.

\subsubsection{L(4,-4,$x$), $x \neq 1$}

$$
\begin{tabular}{|c|c|c|c|c|c|} \hline

{\em Subalgebra } & {\em Complement} & {\em C}  & {$\kappa$}      & {\em Quadratic}   & {\em Petrov} \\
{\lieh}        & {\compm}      & {\em Type}         & {}           & {\em Forms}        & {\em Number}  \\ \hline

$e_1$             & $e_2 + c_1\lieh, e_3 + c_2\lieh, e_4 + c_3\lieh$      &      & 3 & $Q$                &          \\ \hline
$e_2$             & $e_1 + c_1\lieh, e_3 + c_2\lieh, e_4 + c_3\lieh$      &      & 3 & $Q$                &          \\ \hline
$e_3$             & $e_4, e_2, e_1$                                       & S, I & 1 & $D_1$              &          \\ \hline
$e_3 (x=-1)$      & $e_4, e_1 + e_2, -e_1 + e_2$                          & S, I & 1 & $D_1, B$           & 32.11(+) \\ \hline
$e_4$             & $e_1, e_2, e_3 + c\lieh$                              & I    & 3 & $Q$                &          \\ \hline
$e_1 + e_2$       & $e_4 + c_1\lieh, e_3 + c_2\lieh, e_1 + xe_2$          & R    & 2 & $D_1, D_2, Q_{12}$ &          \\ \hline
$e_1+e_2$         & $e_4, e_3 + c\lieh, e_1 - e_2$                        & S    & 2 & $D_1, D_2, Q_{12}$ &          \\
 $(x=-1)$         &                                                       &      &   &                    &          \\ \hline
$e_1 + e_4$       & $e_2, e_3 + c\lieh, e_1$                              & I    & 2 & $D_1, D_2, Q_{12}$ &          \\ \hline
$e_2 + e_4$       & $e_1, e_3 + c\lieh, xe_2$                             & I    & 2 & $D_1, D_2, Q_{12}$ &          \\ \hline
$e_1+e_2+e_4$     & $e_3 + c\lieh, e_1, e_1 + xe_2$                       & I    & 2 & $D_1, D_2, Q_{12}$ &          \\ \hline

\end{tabular}
$$

$$
A_1(t)(v)=\left( \begin{array}{c} y^1+ty^3\\ y^2\\ y^3\\ y^4 \end{array} \right), \;\;
A_2(t)(v)=\left( \begin{array}{c} y^1\\ y^2+txy^3\\ y^3\\ y^4 \end{array} \right), \;\;
O_1(t)(v)=\left( \begin{array}{c} y^1\\ y^2\\ y^3\\ y^4+ty^3 \end{array} \right), \;\;
$$

$$
[S_1(t)(v)]^t=\left( \begin{array}{cccc} \frac{y^1}{t} & y^2 & y^3 & y^4 \end{array} \right), \;\;
[S_2(t)(v)]^t=\left( \begin{array}{cccc} y^1 & \frac{y^2}{t} & y^3 & y^4 \end{array} \right).
$$

The invariants of the Adjoint action are $I = y^3$ and $J = y^4$.

Case 1. \ \  $I \neq 0$: \ \   $\lambda(y^3) \circ O_1(-\frac{y^4}{y^3}) \circ A_2(-\frac{y^2}{xy^3}) \circ A_1(-\frac{y^1}{y^3})(v) = e_3.$

Case 2. \ \  $I = 0, J \neq 0, y^1 \neq 0, y^2 \neq 0$:  \ \  $\lambda(y^4) \circ S_2(\frac{y^2}{y^4}) \circ S_1(\frac{y^1}{y^4})(v) = e_1 + e_2 + e_4.$

Case 3. \ \  $I = 0, J \neq 0, y^1 \neq 0, y^2 = 0$:  \ \  $\lambda(y^4) \circ S_1(\frac{y^1}{y^4})(v) = e_1 + e_4.$

Case 4. \ \  $I = 0, J \neq 0, y^1 = 0, y^2 \neq 0$:  \ \  $\lambda(y^4) \circ S_2(\frac{y^2}{y^4})(v) = e_2 + e_4.$

Case 5. \ \  $I = 0, J \neq 0, y^1 = 0, y^2 = 0$:  \ \  $\lambda(y^4)(v) = e_4.$

Case 6. \ \  $I = 0, J = 0, y^1 \neq 0, y^2 \neq 0$:  \ \  $\lambda(y^2) \circ S_1(\frac{y^1}{y^2})(v) = e_1 + e_2.$

Case 7. \ \  $I = 0, J = 0, y^1 \neq 0, y^2 = 0$:  \ \  $\lambda(y^1)(v) = e_1.$

Case 8. \ \ $I = 0, J = 0, y^1 = 0, y^2 \neq 0$:  \ \   $\lambda(y^2)(v) = e_2.$

\subsubsection{L(4,-4,1)}

$$
\begin{tabular}{|c|c|c|c|c|} \hline

{\em Subalgebra } & {\em Complement} & {\em Complement}  & {$\kappa$}      & {\em Quadratic}   \\
{\lieh}        & {\compm}      & {\em Type}         & {}           & {\em Forms}         \\ \hline

$e_1$             & $e_2 + c_1\lieh, e_3 + c_2\lieh, e_4 + c_3\lieh$      &      & 3 & $Q$                        \\ \hline
$e_3$             & $e_4, e_2, e_1$                                       & S, I & 1 & $D_1$                        \\ \hline
$e_4$             & $e_1, e_2, e_3 + c\lieh$                              & I    & 3 & $Q$                          \\ \hline
$e_1 + e_4$       & $e_2, e_3 + c\lieh, e_1$                              & I    & 2 & $D_1, D_2, Q_{12}$        \\ \hline

\end{tabular}
$$

$$
A_1(t)(v)=\left( \begin{array}{c} y^1+ty^3\\ y^2\\ y^3\\ y^4 \end{array} \right), \;\;
A_2(t)(v)=\left( \begin{array}{c} y^1\\ y^2+ty^3\\ y^3\\ y^4 \end{array} \right), \;\;
O_1(t)(v)=\left( \begin{array}{c} y^1\\ y^2\\ y^3\\ y^4+ty^3 \end{array} \right), \;\;
$$

$$
O_2(t)(v)=\left( \begin{array}{c} \cos(t)y^1+\sin(t)y^2\\ -\sin(t)y^1+\cos(t)y^2\\ y^3\\ y^4 \end{array} \right), \;\;
S_1(t)(v)=\left( \begin{array}{c} y^1/t\\ y^2\\ y^3\\ y^4 \end{array} \right).
$$

The invariants of the Adjoint action are $I = y^3$ and $J = y^4$.

Case 1. \ \  $I \neq 0$:  \ \   $\lambda(y^3) \circ O_1(-\frac{y^4}{y^3}) \circ A_2(-\frac{y^2}{y^3}) \circ A_1(-\frac{y^1}{y^3})(v) = e_3.$

Case 2. \ \  $I = 0, (y^1)^2+(y^2)^2 \neq 0, J \neq 0$:   \ \   

\begin{center}
$\lambda(y^4) \circ S_1\left(\frac{\sqrt{(y^1)^2+(y^2)^2}}{{\rm sign}(y^1)y^4}\right) \circ O_2\left(\arctan\frac{y^2}{y^1}\right)(v) = e_1 + e_4.$
\end{center}

Case 3. \ \  $I = 0, (y^1)^2+(y^2)^2 \neq 0, J = 0$  \ \   

\begin{center}
$\lambda\left(\frac{\sqrt{(y^1)^2+(y^2)^2}}{{\rm sign}(y^1)}\right) \circ O_2\left(\arctan\frac{y^2}{y^1}\right)(v) = e_1.$
\end{center}

Case 4.  \ \ $I = 0, (y^1)^2+(y^2)^2 = 0$: \ \  $\lambda(y^4)(v) = e_4.$

\subsubsection{L(4,-5)}

$$
\begin{tabular}{|c|c|c|c|c|} \hline

{\em Subalgebra } & {\em Complement} & {\em Complement}  & {$\kappa$}      & {\em Quadratic}   \\
{\lieh}        & {\compm}      & {\em Type}         & {}           & {\em Forms}        \\ \hline

$e_1$             & $e_2 + c_1\lieh, e_3 + c_2\lieh, e_4 + c_3\lieh$      &      & 3 & $Q$                      \\ \hline
$e_2$             & $e_4 + c_1\lieh, e_3 + c_2\lieh, e_1 + e_2$           & R    & 2 & $D_1, D_2, Q_{12}$          \\ \hline
$e_3$             & $e_4, e_2, e_1$                                       & S, I & 1 & $D_1$                       \\ \hline
$e_4$             & $e_1, e_2, e_3 + c\lieh$                              & I    & 3 & $Q$                        \\ \hline
$e_1 + e_4$       & $e_3 + c\lieh, e_2, e_1$                              & I    & 2 & $D_1, D_2, Q_{12}$           \\ \hline
$e_2 + e_4$       & $e_1 + e_2 + e_4, e_3 + c\lieh, e_1 + e_2$            & R    & 2 & $D_1, D_2, Q_{12}$           \\ \hline

\end{tabular}
$$

$$
A_1(t)(v)=\left( \begin{array}{c} y^1+ty^3\\ y^2\\ y^3\\ y^4 \end{array} \right), \;
A_2(t)(v)=\left( \begin{array}{c} y^1+ty^3\\ y^2+ty^3\\ y^3\\ y^4 \end{array} \right), \;
A_3(t)(v)=\left( \begin{array}{c} e^{-t}(y^1-ty^2)\\ e^{-t}y^2\\ y^3\\ y^4 \end{array} \right), \;
$$

$$
[O_1(t)(v)]^t=\left( \begin{array}{cccc} y^1& y^2 & y^3 & y^4+ty^3 \end{array} \right), \;\;
[S_4(t)(v)]^t=\left( \begin{array}{cccc} y^1& y^2 & y^3 & \frac{y^4}{t} \end{array} \right).
$$

The invariants of the Adjoint action are $I = y^3$ and $J = y^4$.

Case 1.  \ \  $I \neq 0$:  \ \  $ \lambda(y^3) \circ O_1(-\frac{y^4}{y^3}) \circ A_2(-\frac{y^2}{y^3}) \circ A_1(\frac{y^2-y^1}{y^3})(v) = e_3.$

Case 2. \ \  $I = 0, J \neq 0, y^2 \neq 0$:  \ \   $\lambda\left(y^2e^{-\frac{y^1}{y^2}}\right) \circ S_4\left(\frac{y^4}{y^2}e^{\frac{y^1}{y^2}}\right) \circ A_3(\frac{y^1}{y^2})(v) = e_2 + e_4.$

Case 3. \ \  $I = 0, J \neq 0, y^1 \neq 0, y^2 = 0$:  \ \  $\lambda(y^1) \circ S_4(\frac{y^4}{y^1})(v) = e_1 + e_4.$

Case 4. \ \   $I = 0, J \neq 0, y^1 = 0, y^2 = 0$:  \ \   $\lambda(y^4)(v) = e_4.$

Case 5. \ \   $I = 0, J = 0, y^2 \neq 0$:  \ \   $\lambda\left(y^2e^{-\frac{y^1}{y^2}}\right) \circ A_3(\frac{y^1}{y^2})(v) = e_2.$

Case 6. \ \   $I = 0, J = 0, y^2 = 0:$  \ \   $\lambda(y^1)(v) = e_1.$

\subsubsection{L(4,-6,$x$)}

$$
\begin{tabular}{|c|c|c|c|c|c|} \hline

{\em Subalgebra } & {\em Complement} & {\em C}  & {$\kappa$}      & {\em Quadratic}   & {\em Petrov} \\
{\lieh}        & {\compm}      & {\em Type}         & {}           & {\em Forms}        & {\em Number}  \\ \hline

$e_1$             & $e_4 + c_1\lieh, e_3 + c_2\lieh, xe_1 - e_2$          & R    & 2 & $D_1, D_2, Q_{12}$ &             \\ \hline
$e_1 (x = 0)$     & $e_4, e_3 + c\lieh, -e_2$                             & S    & 2 & $D_1, D_2, Q_{12}$ &             \\ \hline
$e_3$             & $e_4, e_2, e_1$                                       & S, I & 1 & $D_1$              &             \\ \hline
$e_3 (x = 0)$     & $e_4, e_2, e_1$                                       & S, I & 1 & $D_1, R$           &  32.11 ($-$)  \\ \hline
$e_4$             & $e_1, e_2, e_3 + c\lieh$                              & I    & 3 & $Q$                &             \\ \hline
$e_1 + e_4$       & $e_3 + c\lieh, e_1, xe_1 - e_2$                       & I    & 2 & $D_1, D_2, Q_{12}$ &             \\ \hline

\end{tabular}
$$

$$
A_1(t)(v)=\left( \begin{array}{c} y^1+txy^3\\ y^2-ty^3\\ y^3\\ y^4 \end{array} \right), \;\;
A_2(t)(v)=\left( \begin{array}{c} y^1+ty^3\\ y^2+txy^3\\ y^3\\ y^4 \end{array} \right), \;\;
S_4(t)(v)=\left( \begin{array}{c} y^1\\ y^2\\ y^3\\ y^4/t \end{array} \right), \;\;
$$

$$
A_3(t)(v)=\left( \begin{array}{c} e^{-tx}(\cos(t)y^1-\sin(t)y^2)\\ e^{-tx}(\sin(t)y^1+\cos(t)y^2)\\ y^3\\ y^4 \end{array} \right), \;\;
O_1(t)(v)=\left( \begin{array}{c} y^1\\ y^2\\ y^3\\ y^4+ty^3 \end{array} \right).
$$

The invariants of the Adjoint action are $I = y^3$ and $J = y^4$.

Case 1. \ \  $I \neq 0$: \ \  $\lambda(y^3) \circ O_1(-\frac{y^4}{y^3}) \circ A_2(-\frac{y^1+xy^2}{(1+x^2)y^3}) \circ A_1(\frac{y^2-xy^1}{(1+x^2)y^3})(v) = e_3.$

Case 2. \ \  $I = 0, \, J \neq 0, \, H=(y^1)^2+(y^2)^2 \neq 0$
$$\lambda\left(k\right) \circ S_4\left(\frac{{\rm sign}(y^1)y^4}{\sqrt{H}}e^{-\left(\arctan\frac{y^2}{y^1}\right)x}\right) \circ A_3\left(-\arctan\frac{y^2}{y^1}\right)(v) = e_1 + e_4.$$
\hspace{2cm} Note: $\displaystyle k = \frac{\sqrt{H}}{{\rm sign}(y^1)}\exp\left[{\left(\arctan\frac{y^2}{y^1}\right)x}\right]$

\

Case 3. \ \  $I = 0, \, (y^1)^2+(y^2)^2 \neq 0, J = 0$: 
$$\lambda\left(\frac{\sqrt{(y^1)^2+(y^2)^2}}{{\rm sign}(y^1)}e^{\left(\arctan\frac{y^2}{y^1}\right)x}\right) \circ A_3\left(-\arctan\frac{y^2}{y^1}\right)(v) = e_1. $$

Case 4. \ \   $I = 0, \, (y^1)^2+(y^2)^2 = 0$:  \  \  $\lambda(y^4)(v) = e_4$.

\subsubsection{L(4,-7)}

$$
\begin{tabular}{|c|c|c|c|c|c|} \hline

{\em Subalgebra } & {\em Complement} & {\em C}  & {$\kappa$}      & {\em Quadratic}   & {\em Petrov} \\
{\lieh}        & {\compm}      & {\em Type}         & {}           & {\em Forms}        & {\em Number}  \\ \hline

$e_1$                      & $e_2 + c_1\lieh, e_3 + c_2\lieh, e_4 + c_3\lieh$ &    & 2 & $D_1, D_2, Q_{12}$ &             \\ \hline
$e_3$                      & $e_1 + c_1\lieh, e_2 + c_2\lieh, e_4 + c_3\lieh$ &    & 2 & $D_1, D_2, Q_{12}$ &             \\ \hline
$e_4$                      & $e_1, e_2, e_3$                                  & I  & 3 & $Q$                &             \\ \hline
$\frac{1}{2}(e_1 + e_3)$   & $e_4, e_2, \frac{1}{2}(e_1 - e_3)$               & S  & 1 & $D_1, B$           &  32.23 (+)  \\ \hline
$\frac{1}{2}(e_1 - e_3)$   & $e_4, e_2, -\frac{1}{2}(e_1 + e_3)$              & S  & 1 & $D_1, R$           &  32.7       \\ \hline
$e_1 + e_4$                & $\frac{1}{2}e_3, -e_2, e_1$                      & I  & 1 & $D_1, N$           &  32.8       \\ \hline
$\frac{1}{2}(e_1+e_3+e_4)$ & $e_1 + e_3, 2e_2, e_1 - e_3$      & I  & 1 & $D_1, B$           &  32.24 (+)  \\ \hline
$\frac{1}{2}(e_1-e_3+e_4)$ & $e_1 - e_3, -2e_2, e_1 + e_3$      & I  & 1 & $D_1, R$           &  32.24 (--)  \\ \hline

\end{tabular}
$$

$$
A_1(t)(v)=\left( \begin{array}{c} y^1+ty^2-t^2y^3\\ y^2-2ty^3\\ y^3\\ y^4 \end{array} \right), \;\;
A_3(t)(v)=\left( \begin{array}{c} y^1\\ 2ty^1+y^2\\ -t^2y^1-ty^2+y^3\\ y^4 \end{array} \right),
$$

$$
[S_1^3(t)(v)]^t=\left( \begin{array}{cccc} \frac{y^1}{t} & y^2 & ty^3 & y^4 \end{array} \right), \;\;
[S_4(t)(v)]^t=\left( \begin{array}{cccc} y^1 & y^2 & y^3 & \frac{y^4}{t} \end{array} \right).
$$

The invariants of the Adjoint action are $I=4y^1y^3+(y^2)^2$ and $J=y^4$.

Case 1. \ \  $I \neq 0, J \neq 0, y^3 \neq 0$:  \ \   

\hspace{1.5cm} $\lambda(\frac{1}{2}\frac{I}{\sqrt{|I|}}) \circ S_4(2y^4\frac{\sqrt{|I|}}{I}) \circ S_1^3(\frac{1}{2}\frac{\sqrt{|I|}}{y^3}) \circ A_1(\frac{1}{2}\frac{y^2}{y^3})(v) = e_1 + {\rm sign}(I)e_3 + e_4$

Case 2. \ \  $I \neq 0, J \neq 0, y^3=0$:  \ \ 

\hspace{1.5cm} $\lambda(\frac{1}{2}y^2) \circ S_4(2\frac{y^4}{y^2}) \circ S_1^3(2) \circ A_3(-\frac{1}{2}) \circ A_1(\frac{y^2-y^1}{y^2})(v) = e_1 + e_3 + e_4.$

Case 3. \ \  $I \neq 0, J=0, y^3 \neq 0$: \ \  

\hspace{1.5cm} $\lambda(\frac{1}{2}\frac{I}{\sqrt{|I|}}) \circ S_1^3(\frac{1}{2}\frac{\sqrt{|I|}}{y^3}) \circ A_1(\frac{1}{2}\frac{y^2}{y^3})(v) = e_1 + {\rm sign}(I)e_3$.

Case 4. \ \  $I \neq 0, J=0, y^3=0$:  \ \  

\hspace{1.5cm}  $\lambda(\frac{1}{2}y^2) \circ S_1^3(2) \circ A_3(-\frac{1}{2}) \circ A_1(\frac{y^2-y^1}{y^2})(v) = e_1 + e_3$.

Case 5. \ \  $I=0, J \neq 0, y^3 \neq 0$:  \ \ 

\hspace{1.5cm}  $\lambda(y^4) \circ S_1^3(-\frac{y^3}{y^4}) \circ A_3(1) \circ A_1(\frac{y^2-2y^3}{2y^3})(v) = e_1 + e_4.$

Case 6. \ \  $I=0, J \neq 0, y^3=0, y^1 \neq 0:$ \ \  $\lambda(y^1) \circ S_4(\frac{y^4}{y^1})(v) = e_1 + e_4.$

Case 7. \ \  $I=0, J \neq 0, y^3=0, y^1=0:$ \ \  $\lambda(y^4)(v) = e_4.$

Case 8. \ \  $I=0, J=0, y^3 \neq 0$:  \ \  $\lambda(y^3) \circ A_1(\frac{1}{2}\frac{y^2}{y^3})(v) = e_3.$

Case 9. \ \  $I=0, J=0, y^3=0$: \ \  $\lambda(y^1)(v) = e_1$.

\subsubsection{L(4,-8)}

$$
\begin{tabular}{|c|c|c|c|c|c|} \hline

{\em Subalgebra } & {\em Complement}  & {\em Complement}  & {$\kappa$}   & {\em Quadratic}   & {\em Petrov} \\
{\lieh}        & {\compm}       & {\em Type}         & {}        & {\em Forms}        & {\em Number}  \\ \hline

$e_1$             & $e_4, e_3, e_2$   & S                 & 1            & $D_1, R$          &    32.9      \\ \hline
$e_4$             & $e_1, e_2, e_3$   & I                 & 3            & $Q$               &              \\ \hline
$e_1 + e_4$       & $e_1, e_3, e_2$   & I                 & 1            & $D_1, R$          &    32.10     \\ \hline

\end{tabular}
$$

$$
A_1=\left( \begin{array}{c} y^1\\ \cos(t)y^2-\sin(t)y^3\\ \sin(t)y^2+\cos(t)y^3\\ y^4 \end{array} \right), \;
A_3=\left( \begin{array}{c} \cos(t)y^1-\sin(t)y^2\\ \sin(t)y^1+\cos(t)y^2\\ y^3 \end{array} \right), \;
S_4=\left( \begin{array}{c} y^1 \\ y^2 \\ y^3 \\ y^4/t \end{array} \right).
$$

The invariants of the Adjoint action are $I=(y^1)^2+(y^2)^2+(y^3)^2$ and $J=y^4$.

\

\

\

Case 1. \ \  $I \neq 0, J \neq 0$:

\begin{center}
$\lambda\left(\frac{\sqrt{I}}{{\rm sign}(y^1)}\right) \circ S_4\left(\frac{{\rm sign}(y^1)y^4}{\sqrt{I}}\right) \circ A_3(\alpha) \circ A_1\left(-\arctan\frac{y^3}{y^2}\right)(v) =e_1+e_4.$
\end{center}

Case 2.  \ \  $I \neq 0, J = 0$: 

\begin{center}
$\lambda\left(\frac{\sqrt{I}}{{\rm sign}(y^1)}\right) \circ A_3(\alpha) \circ A_1\left(-\arctan\frac{y^3}{y^2}\right)(v) = e_1. $
\end{center}

Case 3. \ \  $I=0, J \neq 0$:  \ \  $\lambda(y^4)(v) = e_4$.

\

Note: In Cases 1 and 2, we use $\alpha=-\arctan\left(\frac{\sqrt{(y^2)^2+(y^3)^2}}{y^1{\rm sign}(y^2)}\right)$

\normalsize

\subsection{Summary of Invariant Metrics}
\label{sec:metric-summary}

\setlength{\parindent}{\defaultparindent}

In Table~\ref{tbl:summary}, we summarize the inequivalent one-dimensional subalgebras of the Lie algebras of three and four dimensions that admit an invariant Lorentz metric.  The notation used is the same as previously.

\renewcommand{\arraystretch}{1.2}
\small

\begin{table}[h]
\caption{\label{tbl:summary} Summary of Invariant Metrics.}
\begin{tabular}{|l|c|c|c|c|c|} \hline
{\em Algebra} & {\em Isotropy} & {\em Complement}  & {\em Com.} & {\em Iso.} & {\em Petrov Number} \\ \hline
L(3,2,$-1$) & $e_3$     & $e_1, e_2$           & S,I   & B       & 30.2       \\ \hline
L(3,4,$0$)  & $e_3$     & $e_2, e_1$           & S,I   & R       & 30.1      \\ \hline
L(3,5)    & $\frac{1}{2}(e_1+e_3)$   & $\frac{1}{2}(e_1-e_3), e_2$   & S   & B   & 30.4    \\ \cline{2-6}
               & $\frac{1}{2}(e_1-e_3)$   & $\frac{1}{2}(e_1+e_3), e_2$   & S   & R   &   \\ \hline
L(3,6)    & $e_1$    &   $e_3, e_2$     &   S   &   R   & 30.6     \\ \hline

L(4,1)      & $e_4$       & $-e_3,e_2,e_1$           & S,I   & N       &  32.12 \\ \hline

L(4,7)      & $e_2+e_3$   & $e_4,e_2,e_1$  & I     & N       &  32.14 ($c=1$)\\ \hline

L(4,8)      & $e_2+e_3$   & $e_4,e_2-e_3,2e_1$       & S     & N       &  32.14 ($c=0$)\\ \cline{2-6}
 
            & $e_4$       & $e_1,e_2+e_3,-e_2+e_3$   & I     & B       &  32.3 \\ \hline

L(4,9,$x$), & $e_2+e_3$   & $\frac{1}{1-x}e_4,\frac{1}{1-x}e_2+\frac{x}{1-x}e_3,$  &   & N   &   32.14 \\ 

$x \neq 1$  &             &  $e_1$ &   &     &     ($c \neq 0, c \neq 1$)    \\ \hline

L(4,10)     & $e_3$       & $e_4,e_2+e_3,e_1$   &       & N       &  32.15 \\ \hline

L(4,11)     & $e_2$       & $-e_4,e_3,-e_1$          & S     & N       &  32.16 ($q=0$)\\ \cline{2-6}

            & $e_4$       & $e_1,e_2,-e_3$           & I     & R       &  32.4\\ \hline

L(4,12,$x$) & $e_2$       & $e_4,xe_2-e_3,e_1$  &       & N       &  32.16 ($q \neq 0$) \\ \hline

L(4,13)     & $e_4$       & $e_3,e_2,e_1$       & I     & R       &  32.6 \\ \hline

L(4,-2)     & $e_2-e_4$   & $e_4, e_1+e_3, -e_1+e_3$         & I     & B   &  32.5 \\ \hline

L(4,-4,-1)  & $e_3$       & $e_4, e_1+e_2, -e_1+e_2$ & S,I   & B       &  32.11 (+) \\ \hline

L(4,-6,0)   & $e_3$       & $e_4,e_2,e_1$            & S,I   & R       &  32.11 (--) \\ \hline

L(4,-7)     & $\frac{1}{2}(e_1+e_3)$   & $e_4, e_2, \frac{1}{2}(e_1-e_3)$   & S   & B   & 32.23 (+) \\ \cline{2-6}

            & $\frac{1}{2}(e_1-e_3)$   & $e_4, e_2, -\frac{1}{2}(e_1+e_3)$  & S   & R   &  32.7 \\ \cline{2-6}

            & $e_1+e_4$                & $\frac{1}{2}e_3, -e_2, e_1$ & I & N &  32.8 \\ \cline{2-6}

            & $\frac{1}{2}(e_1+e_3+e_4)$ & $e_1+e_3, 2e_2, e_1-e_3$  & I & B &  32.24 (+) \\ \cline{2-6}

            & $\frac{1}{2}(e_1-e_3+e_4)$ & $e_1-e_3, -2e_2, e_1+e_3$ & I & R &  32.24 (--) \\ \hline

L(4,-8)     & $e_1$       & $e_4, e_3, e_2$         & S      & R       &  32.9 \\ \cline{2-6}

            & $e_1+e_4$   & $e_1, e_3, e_2$         & I      & R       &  32.10 \\ \hline

\end{tabular}
\end{table}

\normalsize

\section{Final comments}

\subsection*{Conclusions}  
Using the methods of Section~\ref{sec:worksheets}, a total of twenty inequivalent three-dimensional homogeneous spaces were constructed as four-dimensional group actions with one-dimensional isotropy (see Table~\ref{tbl:summary}).  These twenty match up in a one-to-one correspondence with the local group actions given by Petrov; hence, his classification has been verified for this case.  Furthermore, since the invariants are distinct in each case, we have proved Theorem~\ref{thm-petrov}.

For the case of two-dimensional homogeneous spaces, a total of five spaces were constructed as three-dimensional group actions with one-dimensional isotropy.  In this case, it can be seen that Petrov omitted one example.

We observe that five of the $G_4$ actions reduce to  $G_3$ actions.  In these cases, the four-dimensional Lie algebra is decomposable, and has a one-dimensional abelian part.  The action of the isotropy on the three-dimensional part is the same as in the $G_3$ case.  We list these related group actions in Table~\ref{tbl:reductions}.  We will use the Petrov Number to identify the group actions, except for the missing $G_3$ action, which we denote by $M$.

\renewcommand{\arraystretch}{1.2}
\begin{table}[h]
\caption{\label{tbl:reductions} Equivalences between $G_4$ and $G_3$ actions.}
\begin{tabular}{|l|l|c|l|l|} \hline
{\em 4-D Lie Algebra} & 3-D Component & {\em Isotropy}  & $G_4$ & $G_3$ \\ \hline
L(4,-4,-1) & L(3,2,-1)  & $e_3$                                  &  32.11 ($+$)      & 30.2       \\ \hline
L(4,-6,0)   & L(3,4,0)   & $e_3$                                  &  32.11 ($-$)       & 30.1      \\ \hline
L(4,-7)      & L(3,5)       & $\frac{1}{2}(e_1+e_3)$    &  32.23 ($+$)      & 30.4      \\ \cline{3-5}
                  &                   & $\frac{1}{2}(e_1-e_3)$     &  32.7                   & M           \\ \hline
L(4,-8)      & L(3,6)       & $e_1$                                  &  32.9                   & 30.6     \\ \hline
\end{tabular}
\end{table}
\renewcommand{\arraystretch}{1.0}

\subsection*{Group actions of higher dimensions}
In \cite{petrov}, Petrov included group actions of five and six dimensions.  Constructing the group actions as in Section~\ref{sec:worksheets}, by identifying all inequivalent one-dimensional subalgebras of all Lie algebras of dimension five and six, is a daunting task.  There are, for example, 52 indecomposable five-dimensional Lie algebras in Winternitz's list and 27 (non-abelian) decomposable ones.  The author is currently working on an alternate form of verification for these higher dimensional cases.

\subsection*{Acknowledgments}
This paper was developed out of the author's Master's Thesis (submitted in 2002).  The thesis was written under the direction of Ian Anderson, while the author was a student at Utah State University.  The author is indebted to Ian Anderson, Mark Fels, and Charles Torre for their guidance and support during that time.   The author also wishes to express his gratitude to John Stevens.  The list of equivalent 4-D group actions in Table~\ref{tbl:petrov-equiv} was provided by Charles Torre.  Most of the calculations in this paper were done using Ian Anderson's \verb+Maple+ package \verb+DifferentialGeometry+.

\appendix 

\section{The Winternitz Classification}
\label{app:winternitz_tables}

The following classification of Lie algebras is due to P. Winternitz \cite{winternitz-to-appear}.  It is a modification of the well-known classification in \cite{winternitz-1976}.  We note that L(1,1) denotes the one-dimensional abelian Lie algebra.

\renewcommand{\arraystretch}{1}

\footnotesize

\begin{longtable}{|ll|ll|}

\caption{Non-abelian Lie Algebras of Dimensions 2, 3, and 4}
\endfirsthead
\caption{{\em Continued.}}
\endhead
\multicolumn{4}{r}{{\em Continued on next page.}}
\endfoot
\endlastfoot

\hline

L(2,1)
&
\begin{tabular}{l}
\\
$
\begin{array}{c|cccc}
 & e_1 & e_2  \\ \hline
 e_1 & 0 & e_1  \\
 e_2 & -e_1 & 0  \\
\end{array}
$\\
\\
\end{tabular}
&

L(3,1) 
&
\begin{tabular}{l}
\\
$
\begin{array}{c|ccc}
 & e_1 & e_2 & e_3  \\ \cline{1-4}
 e_1 & 0 & 0 & 0  \\
 e_2 & 0 & 0 & e_1  \\
 e_3 & 0 & -e_1 & 0  \\
\end{array}
$\\
\\
\end{tabular}

\\ \hline

L(3,2,$x$)
&
\begin{tabular}{l}
\\
$
\begin{array}{c|ccc}
 & e_1 & e_2 & e_3  \\ \cline{1-4}
 e_1 & 0 & 0 & e_1  \\
 e_2 & 0 & 0 & xe_2  \\
 e_3 & -e_1 & -xe_2 & 0  \\
\end{array}
$ \\
\\
\end{tabular}

&

L(3,3)
&
\begin{tabular}{l}
\\
$
\begin{array}{c|ccc}
 & e_1 & e_2 & e_3  \\ \cline{1-4}
 e_1 & 0 & 0 & e_1  \\
 e_2 & 0 & 0 & e_1+e_2  \\
 e_3 & -e_1 & -e_1-e_2 & 0  \\
\end{array}
$\\
\end{tabular}
\\

\multicolumn{2}{|c|}{$0 < |x| \leq 1$} & & 
\\ \hline

\multicolumn{4}{|l|}
{
L(3,4,$x$)
\ \ \ 
\begin{tabular}{l}
\\
$
\begin{array}{c|ccc}
 & e_1 & e_2 & e_3  \\ \cline{1-4}
 e_1 & 0 & 0 & xe_1-e_2  \\
 e_2 & 0 & 0 & e_1+xe_2  \\
 e_3 & -xe_1+e_2 & -e_1-xe_2 & 0  \\
\end{array}
$\\
\\
\end{tabular}
\ \ \ \ \ \ 
$x \geq 0$
}
\\ \hline

L(3,5)
&
\begin{tabular}{l}
\\
$
\begin{array}{c|ccc}
 & e_1 & e_2 & e_3  \\ \cline{1-4}
 e_1 & 0 & e_1 & -2e_2  \\
 e_2 & -e_1 & 0 & e_3  \\
 e_3 & 2e_2 & -e_3 & 0  \\
\end{array}
$\\
\\
\end{tabular}
&

L(3,6)
&
\begin{tabular}{l}
\\
$
\begin{array}{c|ccc}
 & e_1 & e_2 & e_3  \\ \cline{1-4}
 e_1 & 0 & e_3 & -e_2  \\
 e_2 & -e_3 & 0 & e_1  \\
 e_3 & e_2 & -e_1 & 0  \\
\end{array}
$\\
\\
\end{tabular}

\\ \hline

L(4,1)
&
\begin{tabular}{l}
\\
$
\begin{array}{c|cccc}
 & e_1 & e_2 & e_3  & e_4 \\ \cline{1-5}
 e_1 & 0 & 0 & 0 & 0 \\
 e_2 & 0 & 0 & 0 & e_1 \\
 e_3 & 0 & 0 & 0 & e_2 \\
 e_4 & 0 & -e_1 & -e_2 & 0 \\
\end{array}
$\\
\\
\end{tabular}
&

L(4,3)
& 
\begin{tabular}{l}
\\
$
\begin{array}{c|cccc}
 & e_1 & e_2 & e_3  & e_4 \\ \cline{1-5}
 e_1 & 0 & 0 & 0 & 0 \\
 e_2 & 0 & 0 & 0 & e_1 \\
 e_3 & 0 & 0 & 0 & e_3 \\
 e_4 & 0 & -e_1 & -e_3 & 0 \\
\end{array}
$\\
\\
\end{tabular}
\\ \hline

\multicolumn{4}{|l|}{
L(4,2,$x$,$y$) 
\ \ \
\begin{tabular}{l}
\\

$
\begin{array}{c|cccc}
 & e_1 & e_2 & e_3 & e_4   \\ \cline{1-5}
 e_1 & 0 & 0 & 0 & e_1 \\
 e_2 & 0 & 0 & 0 & xe_2 \\
 e_3 & 0 & 0 & 0 & ye_3 \\
 e_4 & -e_1 & -xe_2 & -ye_3 & 0 \\
\end{array}
$\\
\\
\end{tabular}
\ \ \
$-1 \leq y \leq x \leq 1$, \  $xy \neq 0, x \neq -1$
}
\\ \hline

\multicolumn{4}{|l|}{
L(4,4,$x$)
\ \ \ 
\begin{tabular}{l}
\\

$
\begin{array}{c|cccc}
 & e_1 & e_2 & e_3  & e_4 \\ \cline{1-5}
 e_1 & 0 & 0 & 0 & e_1 \\
 e_2 & 0 & 0 & 0 & e_1 + e_2 \\
 e_3 & 0 & 0 & 0 & xe_3 \\
 e_4 & -e_1 & -e_1-e_2 & -xe_3 & 0 \\
\end{array}
$\\

\\
\end{tabular}
\ \ \
$x\neq 0$
} \\ \hline

\multicolumn{4}{|l|}{
L(4,5,$x$,$y$)
 \ \ \ 
\begin{tabular}{l}
\\
$
\begin{array}{c|cccc}
 & e_1 & e_2 & e_3  & e_4 \\ \cline{1-5}
 e_1 & 0 & 0 & 0 & xe_1 \\
 e_2 & 0 & 0 & 0 & ye_2 - e_3 \\
 e_3 & 0 & 0 & 0 & e_2 + ye_3 \\
 e_4 & -xe_1 & -ye_2+e_3 & -e_2-ye_3 & 0 \\
\end{array}
$\\
\\
\end{tabular}
 \ \ \ \ \ \
$x > 0$
}
\\ \hline

\multicolumn{4}{|l|}{
L(4,6)
\ \ \
\begin{tabular}{l}
\\
$
\begin{array}{c|cccc}
 & e_1 & e_2 & e_3  & e_4 \\ \cline{1-5}
 e_1 & 0 & 0 & 0 & e_1 \\
 e_2 & 0 & 0 & 0 & e_1 + e_2 \\
 e_3 & 0 & 0 & 0 & e_2 + e_3 \\
 e_4 & -e_1 & -e_1-e_2 & -e_2-e_3 & 0 \\
\end{array}
$\\
\\
\end{tabular}
}
\\ \hline

L(4,7)
&
\begin{tabular}{l}
\\
$
\begin{array}{c|cccc}
 & e_1 & e_2 & e_3 & e_4  \\ \cline{1-5}
 e_1 & 0 & 0 & 0 & e_1 \\
 e_2 & 0 & 0 & e_1 & e_2 \\
 e_3 & 0 & -e_1 & 0 & 0 \\
 e_4 & -e_1 & -e_2 & 0 & 0 \\
\end{array}
$\\
\\
\end{tabular}
&

L(4,8)
&
\begin{tabular}{l}
\\
$
\begin{array}{c|cccc}
 & e_1 & e_2 & e_3 & e_4  \\ \cline{1-5}
 e_1 & 0 & 0 & 0 & 0 \\
 e_2 & 0 & 0 & e_1 & e_2 \\
 e_3 & 0 & -e_1 & 0 & -e_3 \\
 e_4 & 0 & -e_2 & e_3 & 0 \\
\end{array}
$\\
\\
\end{tabular}

\\ \hline

\multicolumn{4}{|l|}{
L(4,9,$x$)
 \ \ \ 
\begin{tabular}{l}
\\
$
\begin{array}{c|cccc}
 & e_1 & e_2 & e_3  & e_4 \\ \cline{1-5}
 e_1 & 0 & 0 & 0 & (x+1)e_1 \\
 e_2 & 0 & 0 & e_1 & e_2 \\
 e_3 & 0 & -e_1 & 0 & xe_3 \\
 e_4 & -(x+1)e_1 & -e_2 & -xe_3 & 0 \\
\end{array}
$\\
\\
\end{tabular}
 \ \ \ \ \ \
\begin{tabular}{c}
$-1 < x \leq 1$, \\
\\
$x \neq 0$ \\
\end{tabular}
}
\\ \hline

\multicolumn{4}{|l|}{
L(4,10)
\ \ \ 
\begin{tabular}{l}
\\
$
\begin{array}{c|cccc}
 & e_1 & e_2 & e_3 & e_4  \\ \cline{1-5}
 e_1 & 0 & 0 & 0 & 2e_1 \\
 e_2 & 0 & 0 & e_1 & e_2 \\
 e_3 & 0 & -e_1 & 0 & e_2+e_3 \\
 e_4 & -2e_1 & -e_2 & -e_2-e_3 & 0 \\
\end{array}
$\\
\\
\end{tabular}
}
\\ \hline

L(4,11)
&
\begin{tabular}{l}
\\
$
\begin{array}{c|cccc}
 & e_1 & e_2 & e_3  & e_4 \\ \cline{1-5}
 e_1 & 0 & 0 & 0 & 0 \\
 e_2 & 0 & 0 & e_1 & -e_3 \\
 e_3 & 0 & -e_1 & 0 & e_2 \\
 e_4 & 0 & e_3 & -e_2 & 0 \\
\end{array}
$\\
\\
\end{tabular}

&

L(4,13)
&
\begin{tabular}{l}
\\
$
\begin{array}{c|cccc}
 & e_1 & e_2 & e_3 & e_4  \\ \cline{1-5}
 e_1 & 0 & 0 & e_1 & -e_2 \\
 e_2 & 0 & 0 & e_2 & e_1 \\
 e_3 & -e_1 & -e_2 & 0 & 0 \\
 e_4 & e_2 & -e_1 & 0 & 0 \\
\end{array}
$\\
\\
\end{tabular}

\\ \hline

\multicolumn{4}{|l|}{
L(4,12,$x$)
 \ \ \ 
\begin{tabular}{l}
\\
$
\begin{array}{c|cccc}
 & e_1 & e_2 & e_3  & e_4 \\ \cline{1-5}
 e_1 & 0 & 0 & 0 & 2xe_1 \\
 e_2 & 0 & 0 & e_1 & xe_2-e_3 \\
 e_3 & 0 & -e_1 & 0 & e_2+xe_3 \\
 e_4 & -2xe_1 & -xe_2+e_3 & -e_2-xe_3 & 0 \\
\end{array}
$\\
\\
\end{tabular}
 \ \ \ \ \ \ 
$x > 0$
}
\\ \hline

\hline

L(3,-1)
&
\begin{tabular}{l}
\\
$
\begin{array}{c|ccc}
 & e_1 & e_2 & e_3  \\ \cline{1-4}
 e_1 & 0 & e_1 & 0  \\
 e_2 & -e_1 & 0 & 0  \\
 e_3 & 0 & 0 & 0  \\
\end{array}
$\\
\\
\end{tabular}
&

L(4,-1)
&
\begin{tabular}{l}
\\
$
\begin{array}{c|cccc}
 & e_1 & e_2 & e_3 & e_4  \\ \cline{1-5}
 e_1 & 0 & e_1 & 0 & 0 \\
 e_2 & -e_1 & 0 & 0 & 0 \\
 e_3 & 0 & 0 & 0 & 0 \\
 e_4 & 0 & 0 & 0 & 0 \\
\end{array}
$\\
\\
\end{tabular}

\\

\multicolumn{2}{|c|}{L(2,1) $\oplus$ L(1,1)}
&
\multicolumn{2}{|c|}{L(2,1) $\oplus$ L(1,1) $\oplus$ L(1,1)}

\\ \hline

L(4,-2)
&
\begin{tabular}{l}
\\
$
\begin{array}{c|cccc}
 & e_1 & e_2 & e_3  & e_4 \\ \cline{1-5}
 e_1 & 0 & e_1 & 0 & 0 \\
 e_2 & -e_1 & 0 & 0 & 0 \\
 e_3 & 0 & 0 & 0 & e_3 \\
 e_4 & 0 & 0 & -e_3 & 0 \\
\end{array}
$\\
\\
\end{tabular}
&

L(4,-3)
&
\begin{tabular}{l}
\\
$
\begin{array}{c|cccc}
 & e_1 & e_2 & e_3 & e_4  \\ \cline{1-5}
 e_1 & 0 & 0 & 0 & 0 \\
 e_2 & 0 & 0 & e_1 & 0 \\
 e_3 & 0 & -e_1 & 0 & 0 \\
 e_4 & 0 & 0 & 0 & 0 \\
\end{array}
$\\
\\
\end{tabular}

\\

\multicolumn{2}{|c|}{L(2,1) $\oplus$ L(2,1)}
&
\multicolumn{2}{|c|}{L(3,1) $\oplus$ L(1,1)}

\\ \hline

\multicolumn{4}{|l|}{
L(4,-4,$x$)
\ \ \
\begin{tabular}{l}
\\
$
\begin{array}{c|cccc}
 & e_1 & e_2 & e_3 & e_4  \\ \cline{1-5}
 e_1 & 0 & 0 & e_1 & 0 \\
 e_2 & 0 & 0 & xe_2 & 0 \\
 e_3 & -e_1 & -xe_2 & 0 & 0 \\
 e_4 & 0 & 0 & 0 & 0 \\
\end{array}
$\\
\\
\end{tabular}
\hspace{1.2cm}
\begin{tabular}{c}
L(3,2,$x$) $\oplus$ L(1,1) \\
\\
$0 < |x| \leq 1$
\end{tabular}
}

\\ \hline

\multicolumn{4}{|l|}{
L(4,-5)
\ \ \
\begin{tabular}{l}
\\
$
\begin{array}{c|cccc}
 & e_1 & e_2 & e_3 & e_4  \\ \cline{1-5}
 e_1 & 0 & 0 & e_1 & 0 \\
 e_2 & 0 & 0 & e_1+e_2& 0 \\
 e_3 & -e_1 & -e_1-e_2 & 0 & 0 \\
 e_4 & 0 & 0 & 0 & 0 \\
\end{array}
$\\
\\
\end{tabular}
\hspace{1.0cm}
\begin{tabular}{c}
L(3,3) $\oplus$ L(1,1)
\end{tabular}
}
\\ \hline

\multicolumn{4}{|l|}{
L(4,-6,$x$)
\ \ \ 
\begin{tabular}{l}
\\
$
\begin{array}{c|cccc}
 & e_1 & e_2 & e_3 & e_4  \\ \cline{1-5}
 e_1 & 0 & 0 & xe_1-e_2 & 0 \\
 e_2 & 0 & 0 & e_1+xe_2 & 0 \\
 e_3 & -xe_1+e_2 & -e_1-xe_2 & 0 & 0 \\
 e_4 & 0 & 0 & 0 & 0 \\
\end{array}
$\\
\\
\end{tabular}
\hspace{0.4cm}
\begin{tabular}{c}
L(3,4,$x$) $\oplus$ L(1,1) \\
\\
$x \geq 0$
\end{tabular}

}
\\ \hline

L(4,-7)
&
\begin{tabular}{l}
\\
$
\begin{array}{c|cccc}
 & e_1 & e_2 & e_3 & e_4  \\ \cline{1-5}
 e_1 & 0 & e_1 & -2e_2 & 0 \\
 e_2 & -e_1 & 0 & e_3 & 0 \\
 e_3 & 2e_2 & -e_3 & 0 & 0 \\
 e_4 & 0 & 0 & 0 & 0 \\
\end{array}
$\\
\\
\end{tabular}
&

L(4,-8)
&
\begin{tabular}{l}
\\
$
\begin{array}{c|cccc}
 & e_1 & e_2 & e_3  & e_4 \\ \cline{1-5}
 e_1 & 0 & e_3 & -e_2 & 0 \\
 e_2 & -e_3 & 0 & e_1 & 0 \\
 e_3 & e_2 & -e_1 & 0 & 0 \\
 e_4 & 0 & 0 & 0 & 0 \\
\end{array}
$\\
\\
\end{tabular}
\\

\multicolumn{2}{|c|}{L(3,5) $\oplus$ L(1,1)}
&
\multicolumn{2}{|c|}{L(3,6) $\oplus$ L(1,1)}

\\ \hline

\end{longtable}

\renewcommand{\arraystretch}{1.0}


\begin{thebibliography}{10}

\bibitem{boothby}
{\sc W.~M. Boothby}, {\em An Introduction to Differentiable Manifolds and
  Riemannian Geometry}, San Diego, California: Academic Press, second~ed.,
  1986.

\bibitem{mylorentz}
{\sc A.~Bowers}, {\em An algebraic construction of {L}orentz homogeneous spaces
  of low dimension}.
\newblock Submitted.

\bibitem{Calvaruso5}
{\sc G.~Calvaruso}, {\em Einstein-like curvature homogeneous {L}orentzian
  three-manifolds}, Results Math., 55 (2009), pp.~295--310.

\bibitem{Calvaruso4}
{\sc G.~Calvaruso and B.~De~Leo}, {\em Pseudo-symmetric {L}orentzian
  three-manifolds}, Int. J. Geom. Methods Mod. Phys., 6 (2009), pp.~1135--1150.

\bibitem{Calvaruso6}
{\sc G.~Calvaruso and J.~Van~der Veken}, {\em Lorentzian symmetric three-spaces
  and the classification of their parallel surfaces}, Internat. J. Math., 20
  (2009), pp.~1185--1205.

\bibitem{fels-renner}
{\sc M.~E. Fels and A.~G. Renner}, {\em Non-reductive homogeneous
  pseudo-{R}iemannian manifolds of dimension four}, Canad. J. Math., 58 (2006),
  pp.~282--311.

\bibitem{Galaev}
{\sc A.~S. Galaev}, {\em Einstein spacetimes with recurrent lightlike vector
  fields}, arXiv:1002.4540v1 [math.DG],  (2010).

\bibitem{olver-R2}
{\sc A.~Gonz{\'a}lez-L{\'o}pez, N.~Kamran, and P.~J. Olver}, {\em Lie algebras
  of vector fields in the real plane}, Proc. London Math. Soc. (3), 64 (1992),
  pp.~339--368.

\bibitem{Haesen1}
{\sc S.~Haesen and L.~Verstraelen}, {\em Pseudosymmetry collineations}, J.
  Math. Phys., 48 (2007), pp.~102501, 9.

\bibitem{Haouari}
{\sc N.~Haouari, W.~Batat, N.~Rahmani, and S.~Rahmani}, {\em Three-dimensional
  naturally reductive homogeneous {L}orentzian manifolds}, Mediterr. J. Math.,
  5 (2008), pp.~113--131.

\bibitem{Kiosak}
{\sc V.~Kiosak and V.~S. Matveev}, {\em Complete {E}instein metrics are
  geodesically rigid}, Comm. Math. Phys., 289 (2009), pp.~383--400.

\bibitem{koba72}
{\sc S.~Kobayashi}, {\em Transformation Groups in Differential Geometry},
  Springer-Verlag Berlin Heidelberg, 1995.
\newblock Reprint of the 1972 Edition.

\bibitem{koba-nomi}
{\sc S.~Kobayashi and K.~Nomizu}, {\em Foundations of differential geometry.
  {V}ol. {II}}, Interscience Tracts in Pure and Applied Mathematics, No. 15
  Vol. II, Interscience Publishers John Wiley \& Sons, Inc., New
  York-London-Sydney, 1969.

\bibitem{mac}
{\sc M.~MacCallum}, {\em Locally isotropic spacetimes with non-null homogeneous
  hyper-surfaces}, in Essays in general relativity (A festschrift for A.H.
  Taub), F.~Tipler, ed., New York: Academic Press, 1980, pp.~121--138.

\bibitem{Marvan}
{\sc M.~Marvan and O.~Stol{\'{\i}}n}, {\em On local equivalence problem of
  space-times with two orthogonally transitive commuting {K}illing fields}, J.
  Math. Phys., 49 (2008), pp.~022503, 17.

\bibitem{winternitz-1976}
{\sc J.~Patera, R.~T. Sharp, P.~Winternitz, and H.~Zassenhaus}, {\em Invariants
  of real low dimension {L}ie algebras}, J. Mathematical Phys., 17 (1976),
  pp.~986--994.

\bibitem{petrov}
{\sc A.~Z. Petrov}, {\em Einstein Spaces}, New York: Pergamon Press, 1969.
\newblock Translated by R. F. Kelleher.

\bibitem{winternitz-to-appear}
{\sc L.~Snobl and P.~Winternitz}, {\em Classification and identification of
  {L}ie algebras}.
\newblock To be published.

\bibitem{exact-solutions}
{\sc H.~Stephani, D.~Kramer, M.~MacCallum, C.~Hoenselaers, and E.~Herlt}, {\em
  Exact solutions of {E}instein's field equations}, Cambridge Monographs on
  Mathematical Physics, Cambridge University Press, Cambridge, second~ed.,
  2003.

\bibitem{warner}
{\sc F.~W. Warner}, {\em Foundations of Differentiable Manifolds and Lie
  Groups}, New York: Springer-Verlag, 1983.

\end{thebibliography}

\end{document}